\documentclass[fleqn,usenatbib]{mnras}

\usepackage{newtxtext,newtxmath}

\usepackage[T1]{fontenc}

\DeclareRobustCommand{\VAN}[3]{#2}
\let\VANthebibliography\thebibliography
\def\thebibliography{\DeclareRobustCommand{\VAN}[3]{##3}\VANthebibliography}

\usepackage{xspace}
\usepackage{multirow}
\usepackage{caption}

\usepackage{graphicx}	
\usepackage{amsmath}	

\newcommand{\mystar}{KIC\texorpdfstring{\,}{ }10841730\xspace}
\newcommand{\starLyr}{V538\,Lyr\xspace}
\newcommand{\kepler} {\textit{Kepler}\xspace}
\newcommand{\Echelle} {{\'E}chelle\xspace}
\newcommand{\echelle} {{\'e}chelle\xspace}
\newcommand{\echelles} {{\'e}chelles\xspace}
\newcommand{\mesa} {{\sc mesa}\xspace}
\newcommand{\gyre} {{\sc gyre}\xspace}

\newcommand{\Msun}{M$_{\mathrm{\odot}}$\xspace}

\newcommand{\numax} {$\nu_{\mathrm{max}}$\xspace}
\newcommand{\dnu} {$\Delta\nu$\xspace}
\newcommand{\sdnu}[1]{$\delta\nu_{0,#1}$\xspace}
\newcommand{\dnusurf}{$\delta\nu_{\mathrm{surf}}$\xspace}
\newcommand{\epsp}{$\epsilon_{\mathrm{p}}$\xspace}
\newcommand{\epspp}{$\epsilon_{\mathrm{p}}^{\prime}$\xspace}
\newcommand{\epsg}{$\epsilon_{\mathrm{g}}$\xspace}
\newcommand{\dpi} {$\Delta\Pi_{\mathrm{1}}$\xspace}
\newcommand{\dpil} {$\Delta\Pi_{\ell}$\xspace}
\newcommand{\fdnu} {$f_{\Delta\nu}$\xspace}

\newcommand{\Minit}{$M_{\mathrm{initial}}$\xspace}
\newcommand{\fov}{$f_{\mathrm{ov}}$\xspace}
\newcommand{\aMLT}{$\alpha_{\mathrm{MLT}}$\xspace}

\newcommand{\muHz}{\,$\mu$Hz\xspace}

\newcommand{\Teff}{$T_{\mathrm{eff}}$\xspace}
\newcommand{\logg}{$\log g$\xspace}
\newcommand{\vsini}{$v \sin{i} $\xspace}
\newcommand{\MH}{$\mathrm{[M/H]}$\xspace}

\newcommand{\X}{$\chi^2$\xspace}
\newcommand{\XRC}{$\chi^2_\mathrm{RC}$\xspace}
\newcommand{\XRGB}{$\chi^2_\mathrm{RGB}$\xspace}

\usepackage{CJKutf8}
\newcommand{\CNnames}[1]{{\begin{CJK}{UTF8}{gbsn}~(#1)~\end{CJK}}}

\newcommand{\autorefA}[1]{\hyperref[#1]{Appendix~\ref*{#1}}}
\newcommand{\autorefI}{\hyperref[sec:intro]{Introduction}\xspace}

\newcommand{\autorefD}{\hyperref[sec:dataAv]{data availability}\xspace}

\newcommand*{\myeqref}[2][equation~]{%
  \hyperref[{#2}]{#1(\ref*{#2})}%
}
\def\equationautorefname#1#2\null{%
  equation#1(#2\null)%
}

\newif\ifarxiv
\arxivfalse 

\newcommand{\orcidlink}[1]{\protect\href{https://orcid.org/#1}{\textsuperscript{\protect\includegraphics[width=8pt]{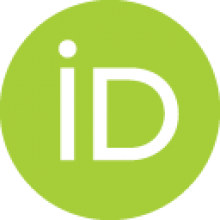}}}}

\title[The Asteroseismic Binary KIC 10841730]{Testing Red Clump Models with the Asteroseismic Binary KIC 10841730}

\author[L. S. Schimak et al.]{Lea S. Schimak\orcidlink{0009-0006-8575-2106},$^{1}$\thanks{E-mail: lea.schimak@sydney.edu.au}
Timothy R. Bedding\orcidlink{0000-0001-5222-4661},$^{1}$ 
Courtney L. Crawford\orcidlink{0000-0002-7654-7438},$^{1}$
Paul G. Beck\orcidlink{0000-0003-4745-2242},$^{2,3}$
\newauthor
Yaguang Li\CNnames{李亚光}\orcidlink{0000-0003-3020-4437},$^{4}$ 
Daniel Huber\orcidlink{0000-0001-8832-4488},$^{1,4}$
Joel Ong\orcidlink{0000-0001-7664-648X},$^{1}$
Benjamin T. Montet\orcidlink{0000-0001-7516-8308},$^{5}$
May Gade Pedersen\orcidlink{0000-0002-7950-0061},$^{1}$
\newauthor
Desmond H. Grossmann\orcidlink{0000-0001-6529-9769},$^{2,3}$
Savita Mathur \orcidlink{0000-0002-0129-0316},$^{2,3}$
Rafael A. García\orcidlink{0000-0002-8854-3776}$^{6}$
\\
$^{1}$Sydney Institute for Astronomy (SIfA), School of Physics, University of Sydney, NSW 2006, Australia\\
$^{2}$Instituto de Astrofísica de Canarias, 38200 La Laguna, Tenerife, Spain\\
$^{3}$Departamento de Astrofísica, Universidad de La Laguna, 38206 La Laguna, Tenerife, Spain\\
$^{4}$Institute for Astronomy, University of Hawai‘i, 2680 Wood-lawn Drive, Honolulu, HI 96822, USA\\
$^{5}$School of Physics, University of New South Wales, Sydney, NSW 2052, Australia\\
$^{6}$University Paris-Saclay, Université Paris Cité, CEA, CNRS, AIM, 91191, Gif-sur-Yvette, France}

\date{Accepted XXX. Received YYY; in original form ZZZ}

\pubyear{2024}

\begin{document}
\label{firstpage}
\pagerange{\pageref{firstpage}--\pageref{lastpage}}
\maketitle

\begin{abstract}

Binaries in which both stars are pulsating are rare but extremely valuable. We present the first study of an asteroseismic binary system consisting of a core helium-burning red clump (RC) star and a red giant branch (RGB) star. The \kepler target KIC 10841730 is a wide binary (period $2917 \pm 8$\,d) that provides ideal conditions to test the accuracy of RC models. While prior studies of RC stars have revealed discrepancies in modelling the period spacings of mixed modes, other model parameters remain largely untested. We perform a detailed modelling analysis using individual mode frequencies and cover a large parameter space in mass, metallicity, He-abundance, mixing length, overshooting, and mass-loss, and we also explore different methods to correct for surface effects.
We find two possible results for the red clump models. One solution requires introducing an unexpected offset of the phase shift in the red clump model, yielding an age consistent with the companion star and current masses of $1.01\pm0.06$ and $1.08\pm0.06$ \,\Msun for the RC and RGB star, respectively. Alternatively, we find that excluding the identification of two questionable radial modes resolves the phase-shift offset issue but results in a higher mass and thus a much younger age for the red clump star, contradicting the age obtained from its companion.
We conclude that uncertainties in red clump models affect not only the g-mode period spacings but also the properties of the p modes.
We show the power of asteroseismic binaries in validating and constraining stellar models and highlight the need for refining red-clump models.
\end{abstract}




\section{Introduction}
\label{sec:intro}
Accurate stellar modelling, constrained by observations, is essential for testing stellar theories.
Asteroseismology, in particular, is a valuable tool to probe the internal structure, complementing analyses based on surface properties alone \citep[see, e.g.,][for a review]{Garcia2019,Aerts2021,Kurtz2022}. Extensive research on modelling low-mass stars has focused on main sequence (MS) and red giant branch (RGB) stars. The red clump (RC) stars, which burn helium after lifting the degeneracy in their core during the He-flash, remain comparatively less explored due to the increased computational demands and uncertainties associated with modelling these stars.

The value of asteroseismic studies of red-giant binary systems has been repeatedly demonstrated \citep{Hon2022,Beck2024,Beck2025}, especially those with eclipses \citep{Hekker2010,Frandsen2013,Beck2014,Rawls2016,Gaulme2016,Brogaard2018,Gaulme2019,Ou2019,Benbakoura2021,Thomsen2022,Brogaard2022,Thomsen2025}. Born from the same gas and dust cloud and being gravitationally bound, they provide additional constraints that a single star cannot provide. However, binaries in which oscillations are detected in both components are still a rarity. Such joint constraints are thus generally limited to stellar clusters, where all members share a common age and composition \citep[e.g.][]{Arentoft2017,McKeever2019,Brogaard2023}. 
\citet{Miglio2014} predicted that we should be able to find only about 200 asteroseismic binaries in the \kepler data, with the majority having two RC components \citep[see also][]{Mazzi2025}. Most of those systems will have their oscillations at similar frequencies, which leads to complex power spectra \citep{Choi2025}. Considering this, the list of binary systems containing evolved stars is even shorter than expected. \citet{LiY2018} found one system containing two subgiants (KIC 7107778). \citet{Beck2018} and \citet{Grossmann2025} studied a subgiant and RGB star (KIC 9163796), and \citet{Murphy2021} studied a $\delta$ Scuti and a secondary clump star (KIC 9773821). A potential binary was found by \citet{Themessl2018} with an RGB and RGB/AGB component (KIC 2568888). \citet{Bell2019} identified 30 lightcurves showing two power excesses. Recently, \citet{Espinoza2025} reported 16 additional \kepler targets with two solar-like power excesses, identifying a few more promising binaries but concluding that most are chance alignments. 

We expanded the list by adding \mystar, the first studied RGB and RC double-oscillating binary system.
The components have similar masses, but the primary is an RC star while its companion is still on the RGB. This system offers an ideal laboratory because the shared age and composition of the two components allow us to test the accuracy of the RC model against the more robust RGB model. This is supported by a spectroscopic analysis to further constrain the models generated by \mesa and \gyre. 

RC stars have been studied in large samples to examine trends and relations between global asteroseismic parameters such as the frequency of maximum oscillation power (\numax), frequency separation of p modes (\dnu), the period spacing of the dipole g modes (\dpi) \citep{Beck2011, Bedding2011, Mosser2012,Vrard2016}, the p-mode phase shift (\epsp) \citep{Kallinger2012} and the behaviour of glitches \citep{Miglio2010,CD2014,Vrard2015,Cunha2015,vrard2022}. The glitches were further analysed using a few selected models that focused on either p modes \citep{CD2014} or g modes \citep{Cunha2015,Matteuzzi2025}. However, a direct comparison of observed individual frequencies with models is generally avoided by the community because of both the computational resources needed and the high model uncertainties. Even secondary clump stars, which ignite helium before the core becomes degenerate and therefore avoid the computationally expensive He-flash, have only been modelled in detail in a few cases \citep[e.g.][Chowhan in prep.]{Murphy2021,Brogaard2023}.
We aim to fill this gap. We focus on a detailed analysis of individual oscillation modes instead of global parameters, to compare model predictions with observations. 

First attempts to match the observed \dpi to models by \citet{Montablan2013} proved promising in testing the core masses of RGB stars, but a discrepancy between models and observations for RC stars became evident, most likely driven by the uncertain description of the growth of the convective core. 
The radiative zone bordering on this convective core acts as the propagation region for the g modes, which means that \dpi is sensitive to the treatment of the convective boundary. The growth of the core can lead to a splitting of its convective zone in models \citep[][]{Eggleton1972}, and it is debated how efficient mixing in these regions would be. Further, when helium is nearly exhausted, so-called breathing pulses could occur \citep[][]{Castellani1985}. The core grows rapidly when fresh He is added, but the He is burned quickly and the core shrinks again. The discrepancy between the observed and modelled \dpi, the uncertain description of the possible splitting of the convection zone and the debated occurrence of the breathing pulses, have all motivated studies to improve and test various descriptions \citep{Montablan2013,Bossini2015,Bossini2017,Constantino2015,Constantino2016,Constantino2017,Blouin2024,Noll2024,Noll2025,Paxton2018,Paxton2019}. Overall, these studies point to the need for a bigger convective core and faster growth rate but there is no clear answer of how to achieve that.
Due to this difficulty, in this work we instead focus on the p modes, which are exclusively sensitive to the outer layers of the star, to obtain another probe to diagnose and maybe improve RC models. 

Since \mystar is a wide binary system, we can assume that mass transfer between the components is negligible, and their evolution has proceeded largely independently. However, mass loss from individual red giants is important, especially at the tip of the RGB, where the star is brightest and largest. The RGB component of \mystar has not yet reached this stage, but the RC component has already passed the He-flash at the tip of the RGB, the phase with the highest expected mass-loss rate. 
This system, in which both stars are oscillating, is therefore an ideal candidate for estimating this mass loss.
The total mass lost during red giant evolution remains an area of active research. Stellar clusters, which, similarly to binaries, share the same age and initial composition, have been a valuable tool. Reimers' law \citep{Reimers1975}, weighted by the scaling factor $\eta$ (see \autoref{sec:model}), is the most commonly used description of this mass loss. Studies measuring the integrated mass loss between RGB and helium-burning stars in clusters lead to discrepant results: investigations of open clusters \citep[e.g.][]{Miglio2012,Stello2016,Handberg2017} typically suggest $\eta<0.2$, whereas observations of globular clusters often imply significantly higher values, with $\eta>0.4$ \citep[e.g.,][]{McDonald2015,Howell2022}. Studies of globular clusters also suggest an increase in mass loss with increasing metallicity \citep[e.g.][]{Talio2020,Howell2024a,Howell2024b}, while \citet{Brogaard2024} and \citet{LiY2025} detected the opposite trend when investigating field stars, which may suggest a more complex dependence.
However, studies like this depend on the accurate determination of the mass of RC stars and, while scaling relations are a useful tool, they are approximations and require model-based corrections \citep[e.g.][]{White2011,Sharma2016}. Modelling individual modes of the stars generally gives more accurate results, but it depends on the quality of the models. In this work, we put the quality of such models to the test. 

\section{Observations} \label{sec:observations}
\subsection{\kepler light curve} \label{sec:lightcurve}

\begin{figure*}
  \includegraphics[width=\linewidth]{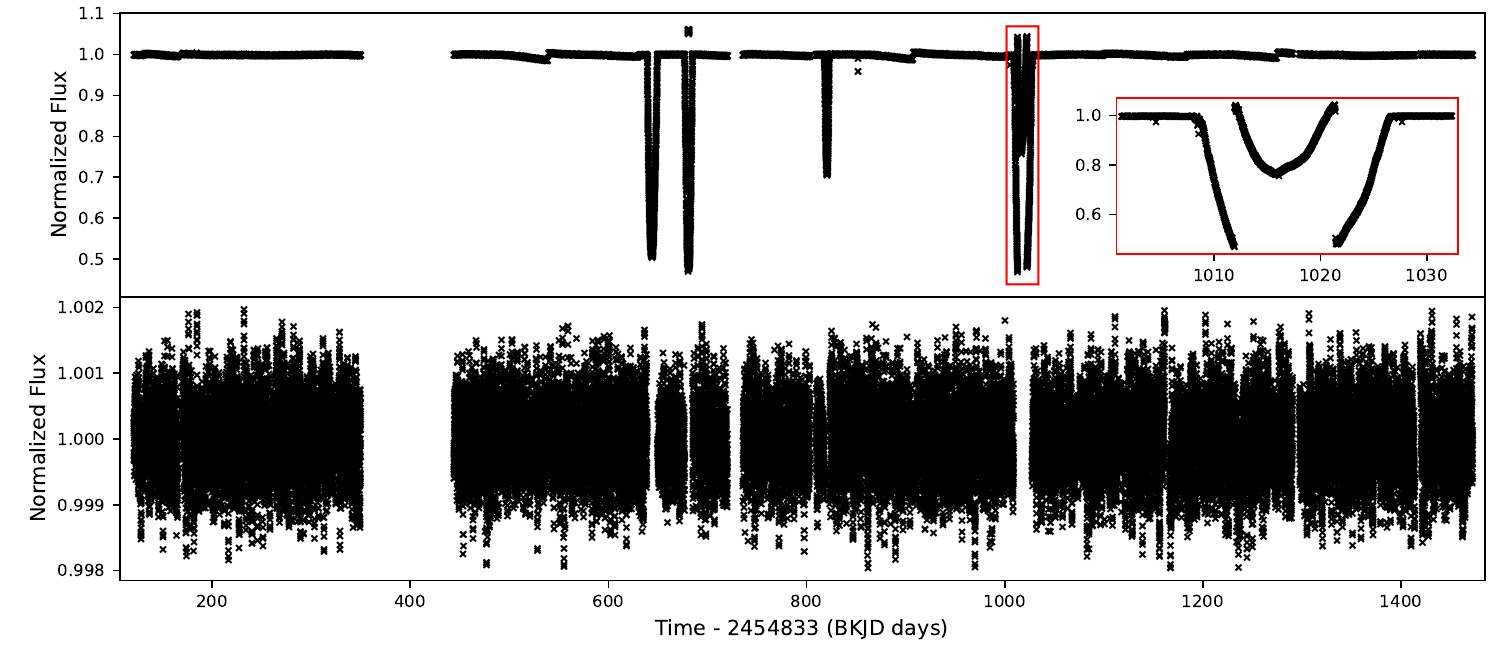}
  \caption{Normalised light curve of \mystar for all available quarters Top: Uncorrected light curve with a zoom on the last dip generated by the overestimated smear correction. The red rectangle marks the zoomed-in area. Bottom: High-pass filtered light curve after we removed the dips and outliers.
  }
  \label{fig:lc_108}
\end{figure*}

\begin{figure}
  \includegraphics[width=\linewidth]{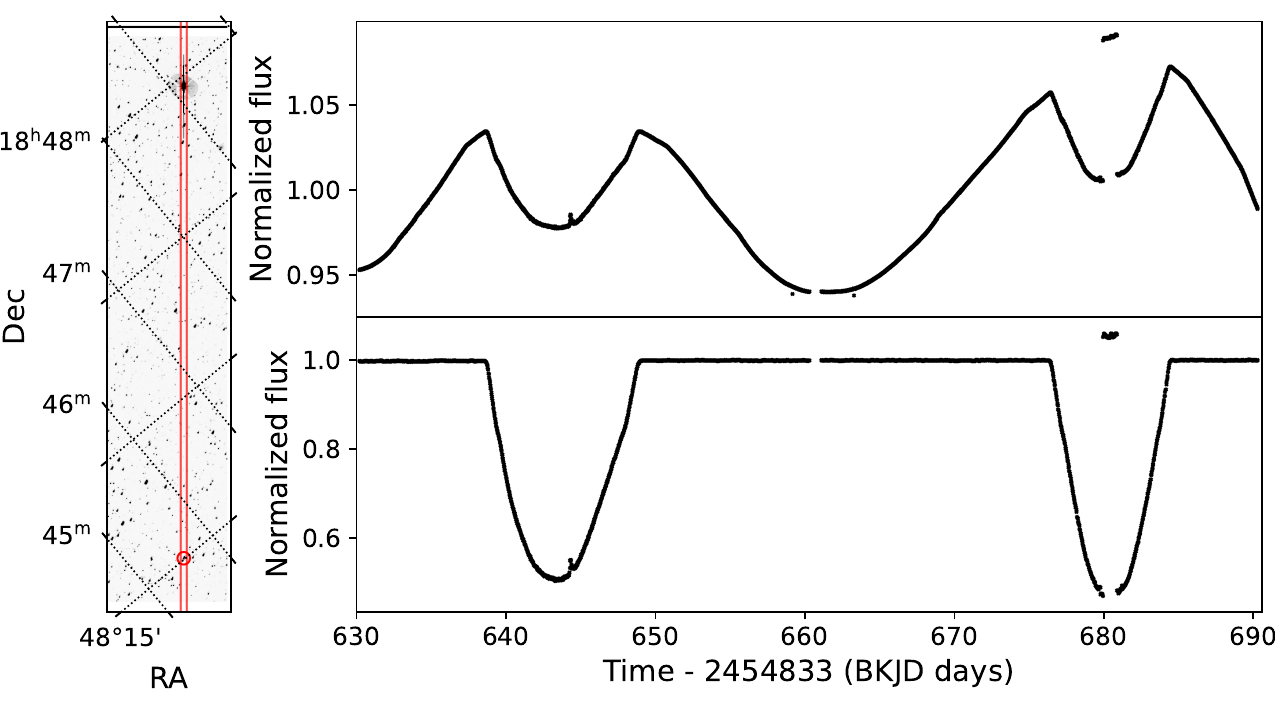}
  \caption{Consequences of the overestimated smear correction. Left: Cutout of the \kepler Full Frame Image (FFI) displaying \mystar, marked by the red circle at the bottom, and the bright long-period variable \starLyr in the same pixel columns, which are highlighted by the thin red lines. Right: Normalised \kepler Simple Aperture Photometry (SAP) light curve for quarter\,7 of \starLyr (top) and \mystar (bottom).}
  \label{fig:FFI_108}
\end{figure}

The binary system \mystar was observed by the \kepler mission \citep{Borucki2010}, yielding nearly four years of high-precision photometry.

We retrieved the light curve from the MAST (Mikulski Archive for Space Telescopes) data archive\footnote{\href{https://mast.stsci.edu/portal/Mashup/Clients/Mast/Portal.html}{https://mast.stsci.edu/portal/Mashup/Clients/Mast/Portal.html}}, presented in the top panel of \autoref{fig:lc_108}. We see four unexpected and irregular dips in flux that cannot be explained by eclipses of the system. It is visible, both in the PDCSAP (Presearch Data Conditioning Simple Aperture Photometry, as shown in the top panel in \autoref{fig:lc_108} and bottom panel in \autoref{fig:FFI_108}) and SAP (Simple Aperture Photometry) flux. Additionally, two of these dips show oddly unphysical behaviour, including a positive flux offset during the minimum of the last observed dip. This results in normalised flux values exceeding unity at the edges of the affected time range. 

A closer look at a monthly Full Frame Image (FFI) corresponding to the orientation of the satellite at the relevant time range, shown in \autoref{fig:FFI_108}, reveals the bright, long-period variable star \starLyr in the same pixel columns as \mystar.
While \starLyr was strongly saturated, its angular separation from \mystar of over 0.9 degrees makes direct contamination highly unlikely. Although \kepler data for \starLyr\ are available for only three quarters, fortunately, one of these overlaps with two of the observed dips in \mystar. This allows us to compare both light curves as shown in \autoref{fig:FFI_108}. Both dips can be found in the light curves of both stars when \starLyr is expected to peak in brightness. We note that the \mystar light curve was only affected when \starLyr was brightest. Interestingly, a small portion of the second dip (around 680 days) appears unaffected, with the flux value consistent with the expected long-period variation of \starLyr. 

To investigate further, we generated a light curve using raw count data instead of the Simple Aperture Photometry (SAP) flux.  This raw light curve shows an uninterrupted long-term variation for \starLyr and no irregularities in the \mystar light curve. We concluded that the dips are because of the smear correction \citep{Quintana2010}. As discussed by \citet{Pope2016,Pope2019},
the \kepler satellite had no shutter, and the CCDs were exposed during readout. This generated a "smear signal" affecting the entire column of the image. Additional pixels were set aside at the end of each column to measure this smear signal and to correct it for affected stars. \starLyr has large brightness variations and was close to the edge of the detector and to the calibration pixels. The saturation signal of the star bled into those pixels, which resulted in an overestimation of the smear correction. This was subtracted from the entire pixel column and produced the dips shown in the light curves. For all further analysis, we removed the affected data points.

To obtain the light curve, we calculated a power spectrum for the target pixel files (TPF) for each quarter. We compared different apertures generated by varying the flux threshold that decides which pixels are included and selected the one resulting in the highest signal-to-noise ratio (SNR). The SNR was estimated by comparing the most prominent oscillation peaks in the power spectra to the white noise. 
Further, we removed long-term trends by applying a high-pass filter, using a Gaussian with a width of 5\,d and excluded a few outlier data points using 2-$\sigma$ sigma-clipping. The final light curve is shown in the bottom panel of \autoref{fig:lc_108}\footnote{The link to the final light curve can be found in the \autorefD section.}.
\subsection{Spectroscopy} \label{sec:spectroscopy}
We observed the system with HERMES \citep[\textit{High Efficiency and Resolution Mercator Echelle Spectrograph,}][]{Raskin2011} and HDS \citep[\textit{High Dispersion Spectrograph,}][]{Noguchi2002,Sato2002} mounted on the Mercator and Subaru telescopes, respectively. A total of 39 HERMES and two HDS spectra were obtained over 12 years. Given the long orbital period of the binary system of nearly 3000\,d (see \autoref{fig:rv}), such a long series of observations is essential to cover the orbit. 
 
The initial data reduction of the raw spectra was performed using the respective instrument pipelines \citep{Raskin2011,Noguchi2002,Sato2002}, including wavelength calibration, cosmic-ray removal and, for the HERMES spectra, merging of the \echelle orders.    
HERMES has a resolution of 85000 from 380 to 900\,nm, but the SNR at the shorter wavelengths was poor. For all further analysis, we only considered wavelengths above 450\,nm. The peak SNR varies from 7 to 35. The two HDS spectra include wavelength measurements from 554\,nm to 685\,nm with a resolution of 100000 and an SNR of about 200. The typical exposure time of all spectra was 30\,min. In a few cases of bad weather conditions and thus bad SNR, two spectra were observed on the same night and then merged.
 
We used the Python framework {\sc iSpec} \citep{Blanco2014,Blanco2019} to perform our spectral analysis. The framework includes various functions to handle different aspects of spectroscopic data analysis and uses the model atmospheres from MARCS (Model Atmospheres with a Radiative and Convective Scheme; \citealt{Gustafsson2008}). 
We used the radiative transfer codes SPECTRUM \citep{Gray1994} and MOOG \citep{Sneden2012}, both implemented within {\sc iSpec}, for spectral fitting.

\subsubsection{Radial velocities} \label{sec:RV}
\begin{figure}
  \includegraphics[width=\linewidth]{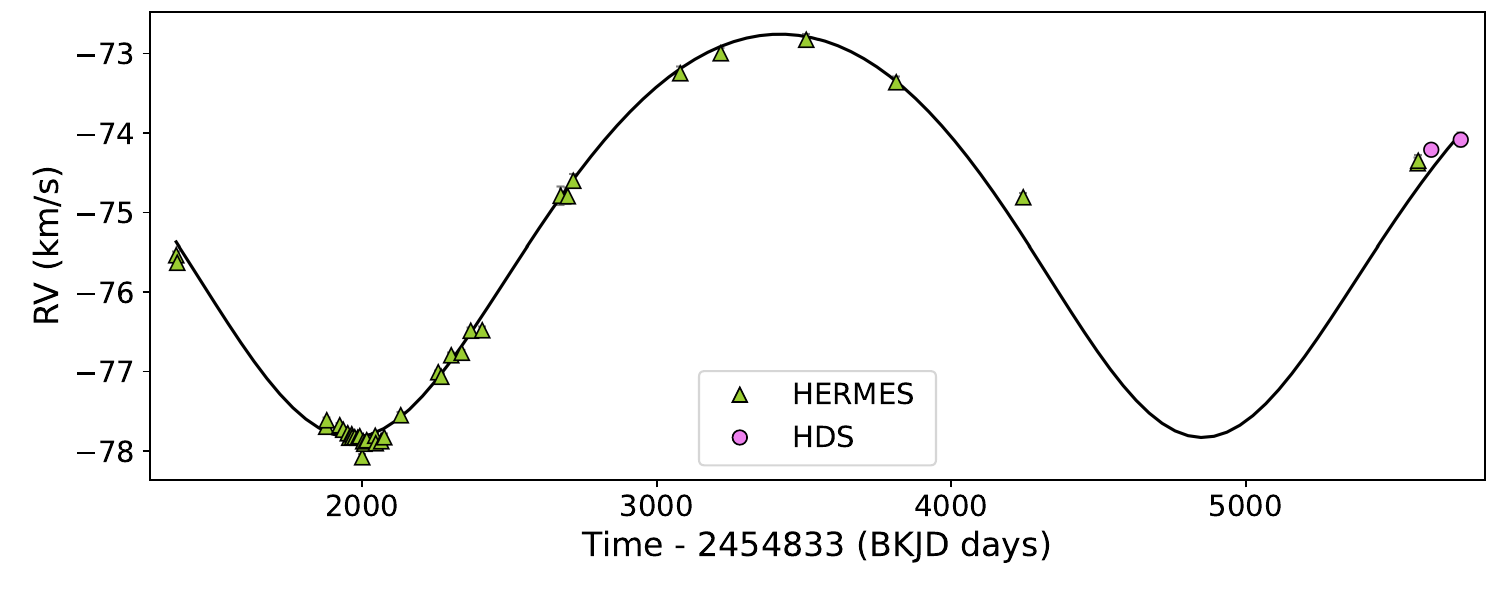}
  \caption{Radial velocities of the absorption line centres of \mystar. The green triangles and pink circles correspond to the points obtained with the HERMES and HDS spectra, respectively. The black line shows the Keplerian fit. For several measurements, the error bars are smaller than the symbol size and therefore not visible.}
  \label{fig:rv}
\end{figure}

We first fitted the continuum and masked out the telluric lines. We then estimated the radial velocities (RVs), shown in \autoref{fig:rv}, by cross-correlating the spectra with a synthetic spectrum of a representative red giant star \citep[][]{Tonry1979}. Next, we fitted a Gaussian to the result of the cross-correlation. We calculated the uncertainty by splitting the spectra into 10 segments and repeating the method to obtain the RVs for each segment.

Our attempts to disentangle the spectra, through the fitting of two Gaussians to a 1D cross-correlation, 2-D cross-correlation \citep[e.g.][]{Mazeh1994}, or the \textit{shift\&add} method \citep{Gonzalez2006}, gave unreliable and inconsistent results. This is not unexpected because of the low SNR and the small difference in RV of the two binary components. The absorption lines of the spectra blend together. 
The RC star dominates the spectra because of its higher luminosity. A very rough estimate based on our disentangling attempts would put the RC star at $\sim 3$ times brighter than the RGB star, which is also supported by examining isochrones (see \autoref{fig:isochrones}).
By estimating the RVs assuming a single-star system, we obtained the RV shift of the line centres, which do not directly correspond to the RVs of the primary star. They slightly underestimate the amplitude of the primary RV curve because the secondary shifts the line centres towards its unknown RV values, depending on its unknown absorption line depth.
 
The uncertainties of the RVs in \autoref{fig:rv} and \autoref{tab:spec_summary} reflect the precision of the measurement. However, the unknown influence of the secondary star, the RGB star, is expected to introduce additional, potentially much larger, systematic uncertainties. Using these RV measurements, we fitted the curve with a Keplerian model using MCMC \citep[Markov chain Monte Carlo,
][]{Foreman2013} and obtained a period of $2917 \pm 8 $\,d. Because we could not disentangle the spectra, we also needed to consider the induced bias on the atmospheric parameters obtained by fitting synthetic spectra in \autoref{sec:specFit}.
With the currently available data, we could not fully exploit the binary nature of the system. Additional, dedicated observations could make disentangling the components possible, providing us with the mass ratio of the two components. Furthermore, future data releases from Gaia \citep[][]{Gaia2023}, including the orbital inclination of the system, may allow us to determine the masses independently of asteroseismology or stellar modelling. This would provide us with additional strong constraints.

\begin{table}   
\caption{Observation times and radial velocities of \mystar}
\label{tab:spec_summary}
\tabcolsep=6.5pt 
\hfill\begin{tabular}{ r r } 
 \hline
 \hline

\multicolumn{2}{c}{HERMES} \\
\multicolumn{1}{c}{BKJD - 2454833 [d]} & 
\multicolumn{1}{c}{RV [km\,s$^{-1}$]} \\
\hline
   1366.45 & -75.52 $\pm$ 0.04 \\
   1369.47 & -75.61 $\pm$ 0.04 \\
   1875.77 & -77.68 $\pm$ 0.02 \\
   1876.72 & -77.60 $\pm$ 0.03 \\
   1921.67 & -77.66 $\pm$ 0.02 \\
   1932.70 & -77.72 $\pm$ 0.03 \\
   1948.55 & -77.76 $\pm$ 0.03 \\
   1953.49 & -77.82 $\pm$ 0.04 \\
   1961.62 & -77.78 $\pm$ 0.03 \\
   1963.62 & -77.81 $\pm$ 0.03 \\
   1972.51 & -77.81 $\pm$ 0.03 \\
   1981.53 & -77.82 $\pm$ 0.06 \\
   1989.50 & -77.80 $\pm$ 0.02 \\
   1996.61 & -78.06 $\pm$ 0.04 \\
   2001.54 & -77.86 $\pm$ 0.02 \\
   2004.53 & -77.89 $\pm$ 0.03 \\
   2010.58 & -77.86 $\pm$ 0.03 \\
   2012.56 & -77.85 $\pm$ 0.04 \\
   2041.44 & -77.80 $\pm$ 0.04 \\
   2042.39 & -77.89 $\pm$ 0.04 \\
   2061.43 & -77.86 $\pm$ 0.04 \\
   2071.44 & -77.81 $\pm$ 0.04 \\
   2128.33 & -77.54 $\pm$ 0.04 \\
   2254.71 & -77.00 $\pm$ 0.02 \\
   2265.71 & -77.05 $\pm$ 0.03 \\
   2299.65 & -76.78 $\pm$ 0.03 \\
   2335.47 & -76.75 $\pm$ 0.03 \\
   2365.60 & -76.47 $\pm$ 0.04 \\
   2404.47 & -76.47 $\pm$ 0.04 \\
   2671.63 & -74.78 $\pm$ 0.12 \\
   2695.52 & -74.79 $\pm$ 0.10 \\
   2713.67 & -74.59 $\pm$ 0.08 \\
   3076.61 & -73.24 $\pm$ 0.08 \\
   3214.36 & -72.99 $\pm$ 0.06 \\
   3505.49 & -72.81 $\pm$ 0.07 \\
   3808.51 & -73.35 $\pm$ 0.07 \\
   4239.47 & -74.80 $\pm$ 0.05 \\
   5579.74 & -74.37 $\pm$ 0.07 \\
   5580.69 & -74.34 $\pm$ 0.07 \\
 \hline
\multicolumn{2}{c}{HDS} \\
\hline
   5625.95 & -74.20 $\pm$ 0.07 \\
   5724.81 & -74.07 $\pm$ 0.09 \\
 \hline
\end{tabular} \hfill~
\end{table}

\subsubsection{Atmospheric Parameters} \label{sec:specFit}
\begin{figure}
  \includegraphics[width=\linewidth]{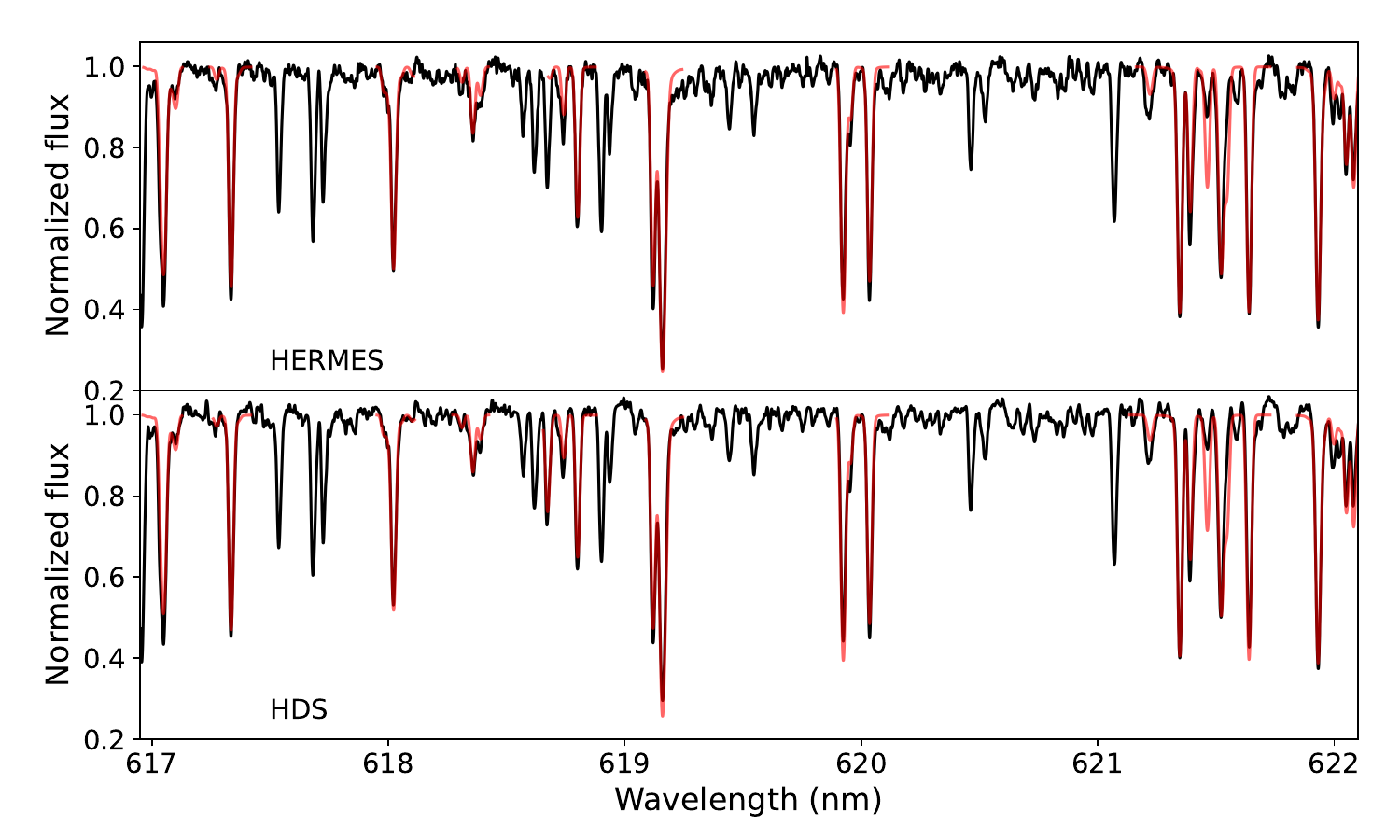}
  \caption{Combined spectrum of \mystar observed with HERMES (top panel) and HDS (bottom panel) with the SPECTRUM fitted synthetic spectrum in red (see \autoref{sec:specFit}).}
  \label{fig:spectra}
\end{figure}

To enhance the SNR, we combined all the spectra obtained with the same instrument by shifting them to the rest frame of the primary, the RC star. 
We fitted synthetic spectra to the combined spectra to estimate the effective temperature (\Teff), the metallicity (\MH), surface gravity (\logg) and the rotational line broadening (\vsini). In our case, this also includes the line broadening due to the contribution from the secondary. Alternatively, \logg could also be constrained using the asteroseismic scaling relations. We decided against this approach because of the high uncertainties due to the spectra not being disentangled and the secondary contaminating the true \logg, depending on the orbital position. 
We selected only neutral and singly-ionised iron lines and generated synthetic spectra with both radiative transfer codes SPECTRUM \citep{Gray1994} and MOOG \citep{Sneden2012}. The comparison of two different codes provided an estimate of systematic errors. We generated the line list, which we used for fitting, by comparing the spectra to an atomic line list taken from the Vienna Atomic Line Database \citep[VALD;][]{Kupka2011}. We defined the square root of the inverse of the continuum fitted flux as the uncertainty of the spectra, which we used to weight the spectral synthesis fit.
We compare the results for the combined spectra from the different instruments and different codes to fit in \autoref{tab:specFit}. The corresponding fit is shown in \autoref{fig:spectra}. The uncertainties reported by {\sc iSpec} were unreasonably small and, together with the fact that we were unable to disentangle these spectra, we instead adopted an uncertainty of $300$\,K for the temperature and $0.2$ for the metallicity.
We used the results from HDS, due to the higher SNR of the individual spectra. Thus, our final spectroscopic results are \Teff$ =4620 \pm 300$\,K for the RC star and \MH$=-0.06\pm0.20$ for both components. 

\begin{table}
\caption{Synthetic spectra fit results of the atmospheric parameters. M marks the results obtained with MOOG and S with SPECTRUM}
\label{tab:specFit}
\tabcolsep=4pt 
\begin{tabular}{c | r r | r r | r r } 
 \hline
 \hline
  & 
\multicolumn{2}{c}{\Teff [K]} & 
\multicolumn{2}{c}{\MH} 
\\

&
\multicolumn{1}{c}{M} & \multicolumn{1}{c}{S} & 
\multicolumn{1}{c}{M} & \multicolumn{1}{c}{S} 
\\

 \hline
  \begin{tabular}{c} HDS \end{tabular}    
  & 4575 $\pm$ 300 & 4620 $\pm$ 300 
  & -0.04 $\pm$ 0.20 & -0.06 $\pm$ 0.20 
  \\
  \begin{tabular}{c} HERMES \end{tabular} 
  & 4624 $\pm$ 300 & 4669 $\pm$ 300 
  & -0.02 $\pm$ 0.20 & -0.01 $\pm$ 0.20 
  \\
\hline
\end{tabular} 
\end{table}

\section{Oscillation frequencies} \label{sec:asteroseismology}
We calculated the power spectral density (PSD) of the light curve using a Lomb-Scargle periodogram \citep{Lomb1976,Scargle1982}. Low-frequency signals, caused by stellar activity, rotation, or various instrumental effects, were removed using a high-pass filter described in \autoref{sec:lightcurve}.

\begin{figure*}
  \includegraphics[width=\linewidth]{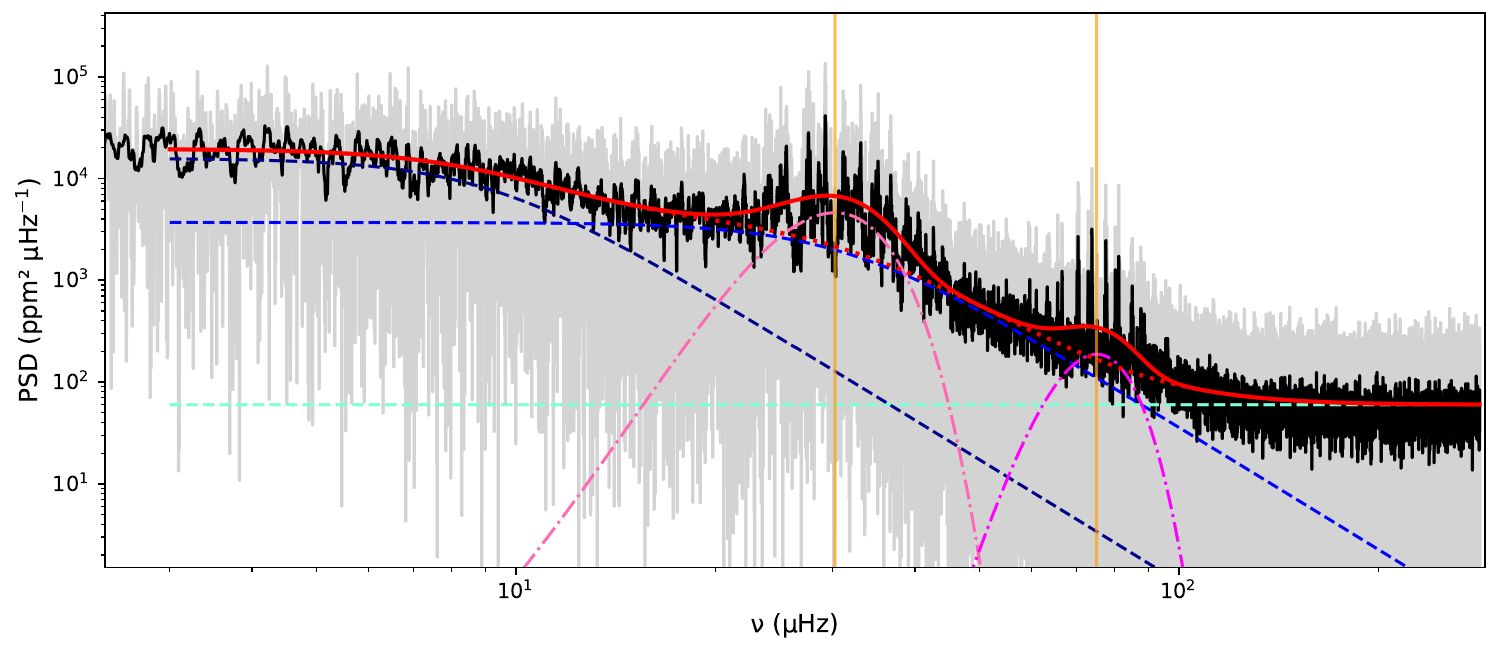}
  \caption{Power spectrum density (PSD) of \mystar displaying the different components of the fit. The grey and black lines represent the PSD before and after smoothing and rebinning. The pink dot-dashed lines correspond to the Gaussians of the solar-like oscillations, the two blue dashed lines are the super-Lorenzians, and the horizontal turquoise dashed line is the photon noise. The solid red line depicts the total fit, and the dotted red line is the background without the solar-like oscillations. The golden vertical lines mark the resulting \numax of both stars.}
  \label{fig:backgroundFit}
\end{figure*}

We modelled the PSD, following \citet{Kallinger2014}, as the sum of two super-Lorentzian functions induced by granulation with an exponent of 4, two Gaussians representing the solar-like oscillations of both stars, a constant offset for the photon noise, and a factor that defines the attenuation of the signal close to the Nyquist frequency. We fitted this model to the smoothed and rebinned PSD using a nonlinear least-squares approach and show the result in \autoref{fig:backgroundFit}. The centres of both Gaussians provided our measurements of \numax found in \autoref{tab:globalSeismic}. 

We describe the individual oscillation modes using the spherical harmonics, the angular degree $\ell$ and radial order $n$. The large frequency separation \dnu corresponds to the difference between modes with the same $\ell$ but consecutive $n$ \citep{Tassoul1980}. We split the power spectrum into equal parts, of length \dnu, and stacked them on top of each other to obtain the \echelle diagram \citep{Grec1983}, shown in panels (b) and (f) in \autoref{fig:streched}. The frequencies of the oscillation modes $\nu_{n,\ell}$ (with the radial order $n$ and the angular degree $\ell$) can be described as 
\begin{equation} \label{eq:epsp}
\nu_{n,\ell} = \Delta \nu (n + \ell/2 + \epsilon_{p}(\nu_{n,\ell}))+\delta \nu_{0,\ell}(\nu_{n,\ell}).
\end{equation} 
The phase shift (\epsp) describes the horizontal offset from zero in the \echelle diagram of the radial p modes, and \sdnu{\ell} is the small frequency separation between the radial modes ($\ell=0$) and modes of the angular degree $\ell$. 
Using the \echelle diagram, we adjusted \dnu to align the radial modes vertically.
The values of \numax and \dnu for both stars are shown in \autoref{tab:globalSeismic}. The uncertainties were determined by splitting up the light curve into seven sections, each consisting of two \kepler quarters, and refitting the background and the power excesses, and remeasuring \dnu for each section. The standard error was adopted as the uncertainty. 

\begin{table}
\caption{Global asteroseismic parameters for both components of \mystar.}
\label{tab:globalSeismic}
\tabcolsep=6.5pt 
\hfill\begin{tabular}{l | r r r} 
 \hline
 \hline
  & 
\multicolumn{1}{c}{\numax} & 
\multicolumn{1}{c}{\dnu} & 
\multicolumn{1}{c}{\dpi}  \\
&
\multicolumn{1}{c}{[\muHz]} & 
\multicolumn{1}{c}{[\muHz]} &
\multicolumn{1}{c}{[s]} 
\\ 
 \hline
  RC  & 30.0 $\pm$ 0.3 & 3.69 $\pm$ 0.02 & 236.9 $\pm$ 0.5\\
  RGB & 74.6 $\pm$ 0.9 & 7.20 $\pm$ 0.03 & 72.6  $\pm$ 0.2\\
 \hline 
\end{tabular} \hfill~
\end{table} 

\subsection{Mode selection}
\label{sec:get_modes}
The power excesses of the oscillations of the two stars are well separated in frequency and do not overlap (see \autoref{fig:backgroundFit}). We could therefore treat the asteroseismic analysis and mode identification of both stars independently, as if they were single stars.
We estimated the frequencies of the radial ($\ell=0$) and quadrupole ($\ell=2$) modes by selecting the frequency range of each relevant peak in the power density spectrum manually and then fitting a Lorentzian using a nonlinear least-squares approach. We scaled the uncertainties by dividing the peak SNR by the width of the fitted Lorentzians. This approach results in very small uncertainties for the strongest modes, in some cases comparable to the systematic frequency shifts caused by the line-of-sight Doppler velocity of the star \citep{Davies2014}. We note that these uncertainties are used primarily as relative weights in our model comparison, and a different treatment would be required to obtain realistic absolute frequency uncertainties. The Doppler shifts themselves are negligible in our analysis, being far smaller than other systematic effects, such as surface terms (see \autoref{sec:surfEff_ball}). The dipole ($\ell=1$) modes were fitted differently. For red giant stars around 1\,\Msun we expect mixed modes, which result from coupling between the acoustic, pressure, or p modes, in the envelope, and the buoyancy, gravity, or g modes in the interior radiation zone. While p modes are approximately equally spaced in frequency, the dipole g modes are approximately equally spaced in period with the $\ell$-dependent period spacing (\dpil) \citep{Tassoul1980}. 
 
\subsubsection{Period spacing}
The asymptotic period spacing of g modes is given by 
\begin{equation}
\Delta \Pi_{\ell}=\frac{2 \pi^2}{\sqrt{\ell(\ell+1)}}\left(\int_{r_1}^{r_2} \frac{N}{r} d r\right)^{-1},
\end{equation}
where $N$ is the the Brunt V\"asi\"al\"a or buoyancy frequency, indicating the highest possible frequency of g modes \citep{Tassoul1980}. In red giants, the high-density helium core and the steep composition gradient left by the hydrogen-burning shell lead to a high $N$. This allows the g modes to propagate, and what defines \dpil. In RC stars, convective mixing replaces part of the stratified radiative zone, and thus the size of the convective core directly impacts \dpi. 

In the case of mixed modes, the observed period spacing is affected by the p modes and is smaller when their frequency is close to that of a pure p mode (referred to as a $\pi$ mode). The $\pi$ modes indicate the frequency at which we would observe a p mode if it were not coupled with a g mode \citep{Aizenman1977}. We calculated the stretched period $\tau(\nu)$ to remove this distortion caused by the coupling and to identify the pure g, or $\gamma$ modes \citep{Mosser2015streched,Ong2023}: 
\begin{equation}  
\tau(\nu) = \frac{1}{\nu}+\frac{\Delta \Pi_l}{\pi} \arctan \left(\frac{q(\nu)}{\tan \Theta_p(\nu)}\right).
\end{equation}
Here $\nu$ is the mode frequency and $q$ is the coupling strength, which describes how strongly g modes couple with p modes. The coupling is generally stronger for red clump stars than RGB stars \citep{Mosser2017,Rossem2024,Lier2025}, which explains the greater spread of detected mixed modes (see panels (b) and (f) in \autoref{fig:streched}). Additionally, we need $\Theta_p(\nu)$, which is defined as
\begin{equation}\label{eq:thetap}
\Theta_p(\nu)=\pi\left(\frac{\nu}{\Delta \nu}
-\epsilon_p^{\prime}(\nu)\right).
\end{equation}
Here 
\begin{equation}\label{eq:epprime}
\epsilon_p^{\prime}(\nu) = \ell/2+\delta \nu_{0,\ell}(\nu)+\epsilon_p(\nu),
\end{equation}
which can also be obtained by calculating $\epsilon_p^{\prime} = (\nu \mod \Delta\nu)/\Delta\nu$, to avoid the need to estimate \sdnu{l}.

The stretched period \echelle diagrams are shown in panels (c) and (g) of \autoref{fig:streched}. We estimated \dpi by manually aligning the modes vertically while simultaneously adjusting the coupling factor~$q$ and~\epspp.
The width of the evanescent zone between the regions where p and g modes propagate depends on frequency and, consequently, so does~$q$ \citep{Rossem2024}. We therefore allowed a different $q$ for each radial order. While allowing a frequency dependence of $q$ was not strictly required for an accurate estimation of \dpi, it noticeably improved the vertical alignment of the stretched mixed modes and significantly improved all further analysis in \autoref{sec:stretchedFreq} to obtain the $\pi$ modes.
The final values for \dpi and $q$ were determined by using both the stretched period and frequency \echelles together (see \autoref{sec:stretchedFreq}) by minimising the standard deviation of the stretched mixed modes for each radial order \citep{Jiang2020}. We estimated the uncertainties of \dpi by varying the relevant parameters until we could no longer align the modes. 
The values for $q$ can be found in \autoref{tab:modes} together with all identified modes. Interestingly, we found $q$ to be increasing with frequency for both stars, opposite to the findings of \citet{Rossem2024}. However, when inspecting our RC models, we found that the width of the evanescent zone decreases with frequency. The coupling is stronger for a smaller evanescent zone, and thus our models agree with our observations.

The period spacing, \dpi, probes the core of the star and thus depends on the evolutionary state. The values support our identification of the primary as a helium-burning red clump star (\dpi of $236.9\pm0.5$\,s) and the secondary as a red giant branch star (\dpi= $72.6\pm0.2$\,s) (see \citet{Bedding2011,Mosser2012}). However, for the RGB star we could find at least two additional values for \dpi, besides the one used in \autoref{fig:streched} that align the dipole modes vertically. A similar case where more than one solution for \dpi was found is discussed by \citet{Buysschaert2016}. While these values of \dpi $=68.5 \pm 0.2$\,s and $77.1\pm 0.3$\,s give a higher variance and more pronounced glitches, it is possible that one of these two values for \dpi could be the correct one. This is because we do not detect all the mixed modes, only those with the most p-mode character. Hence, solutions with one more or one fewer mode (thus changing the radial order $n_g$) in between the more p-like modes are feasible. On the other hand, no matter which of the three possible solutions we choose for \dpi, the resulting $\pi$ modes (calculated as described below) are not affected.
In \autoref{fig:streched} panel (g), we can see a pronounced glitch between the two $\gamma$ modes with the lowest frequency, which may be a valuable characteristic to further constrain the inner structure of clump stars and could be investigated further in a future project.
 
\subsubsection{$\pi$ modes}
\label{sec:stretchedFreq}
Because of the following complications, we decided against using the mixed modes to constrain our models. The surface correction (see \autoref{sec:surfEff_ball}) only affects p modes, because g modes do not propagate near the surface. This means that mixed modes are only partly affected, and the required correction depends only on the p-mode component \citep{Ball2018}. The method developed by \citet{Ong2021_1} considers this, but is expensive computationally. As the modelling was already computationally demanding because we had to evolve the star over the He-flash, we decided to use a different method. In addition, RC stars are known to be challenging to model due to the complex convective He burning core boundary, which is reflected by the difficulty in matching the observed to the modelled \dpi \citep{Montablan2013,Constantino2015,Bossini2015}. This in turn also influences the frequencies of the mixed modes (see \autoref{sec:intro} for further details).
Therefore, we decided only to consider the decoupled, pure p-mode component (the $\pi$ modes). This way, we could use the surface correction as intended, and did not need to calculate the g modes in the models. Instead of comparing individual g-mode frequencies, we relied on \dpi for the RGB star and did not consider it for the RC star.

\begin{figure*}
  \includegraphics[width=\linewidth]{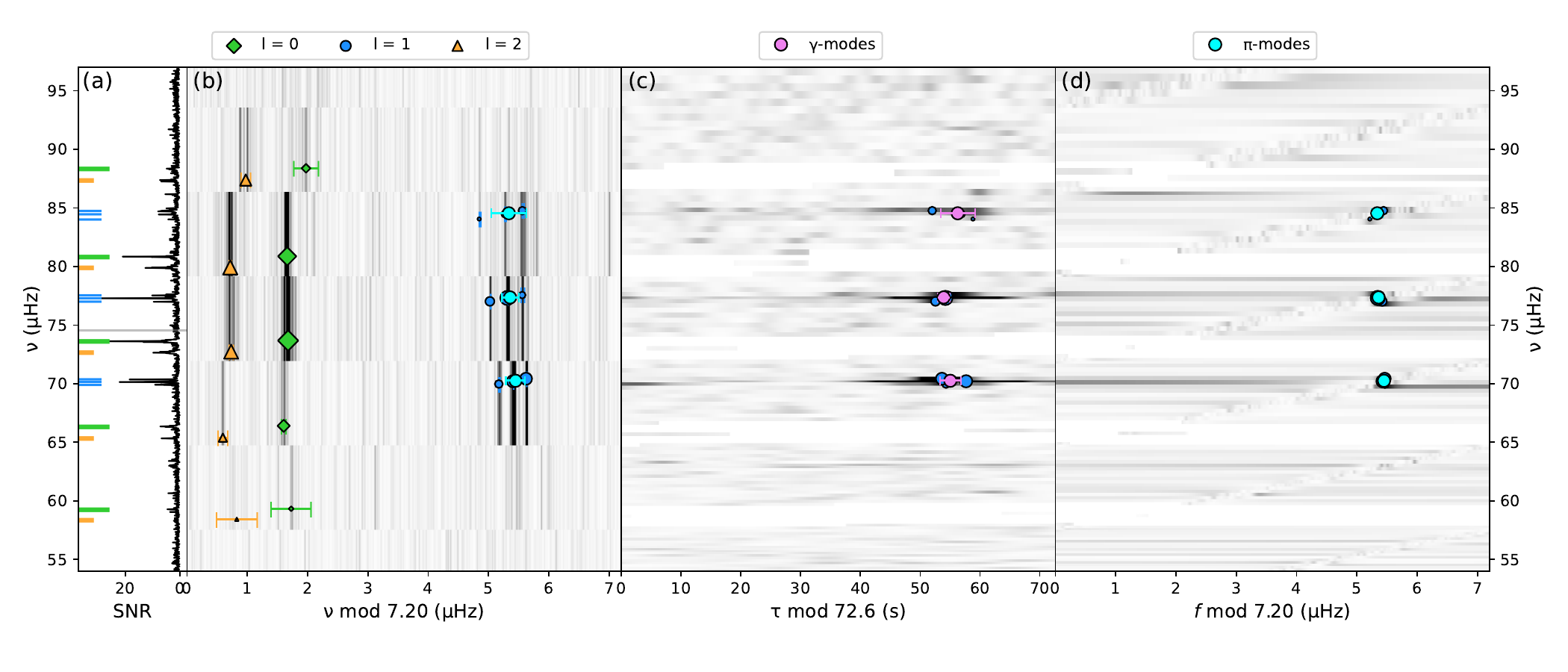}
  \includegraphics[width=\linewidth,trim={0 0 0 1.5cm},clip]{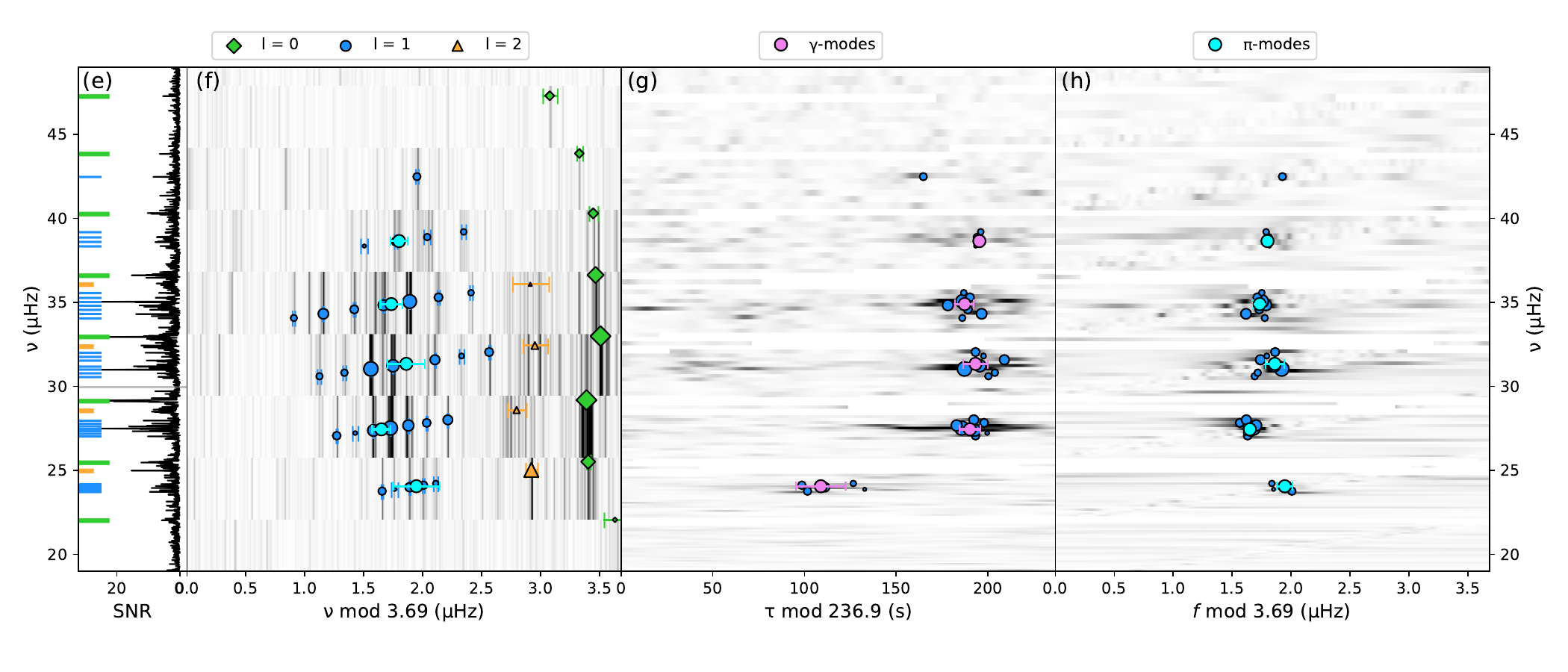}
  \caption{Oscillations of the RGB star (top) and RC star (bottom) of the binary system \mystar. The panels (a) and (e) show the background-corrected power density spectra; (b) and (f) show the frequency \echelles, (c) and (g) show the stretched period \echelles and (d) and (h) show the stretched frequency \echelles. The green diamonds, blue circles, and red triangles represent the $\ell=0, 1, 2$ modes, and the cyan circles show the $\pi$ modes for the frequency \echelles, and the violet circles the average of the $\gamma$ modes for the stretched period \echelles. For clarity, in the stretched \echelles the $\ell=0$ and $2$ modes were masked out to emphasize the $\ell=1$ modes.} 
  \label{fig:streched}
\end{figure*}

For the models, we calculated the $\pi$ modes following \citet{Ong2021}. To estimate them for the observations, we used the stretched frequency $f(\nu)$ described by \citet{LiY2024_conf}, which follows the same principle as the stretched period but for $\pi$ modes instead of $\gamma$ modes. The diagram can be seen in panels (d) and (h) in \autoref{fig:streched}. We calculated it using
\begin{equation}
f(\nu)=\nu-\frac{\Delta \nu}{\pi} \arctan \left(\frac{q\left(\nu\right)}{\tan \Theta_g(\nu) }\right)
\end{equation}
with
\begin{equation}
\Theta_g(\nu) = \pi \left(\epsilon_g(\nu) - \frac{1}{\nu \Delta \Pi}\right).
\end{equation}
Here, \epsg is the phase shift that represents the offset of the pure g modes in a period \echelle, or in this case, the offset of the $\gamma-modes$ in the stretched period \echelle.  
Both \epspp and \epsg were determined by taking the weighted average of the dipole modes for each $n$ in the respective stretched diagrams. Iteratively, we recalculated the stretched period and frequency with the updated values for \epspp and \epsg. To avoid divergence, it was essential to choose suitable initial guesses for \epspp. 
With the final \epspp we obtained the $\pi$ modes, which we listed in \autoref{tab:modes}. 

\begin{table}
\caption{Oscillation mode frequencies sorted by angular degree and coupling strength of both components of \mystar.}
\label{tab:modes}
\tabcolsep=3.5pt 
\hfill\resizebox{\columnwidth}{!}{
\begin{tabular}{r r r r r} 
 \hline
 \hline
 \multicolumn{1}{c}{$\ell=0$} & \multicolumn{1}{c}{$\ell=2$} & \multicolumn{2}{c}{$\ell=1$} & \multicolumn{1}{c}{$q$} \\
 \multicolumn{1}{c}{[\muHz]} & \multicolumn{1}{c}{[\muHz]} & \multicolumn{2}{c}{[\muHz]} & \multicolumn{1}{c}{-}
 \\
  &         & \multicolumn{1}{c}{mixed} &  \multicolumn{1}{c}{$\pi$ modes} & \\
\hline
\multicolumn{5}{c}{RC} \\
\hline
22.057 $\pm$  0.096  &                      &  23.766 $\pm$  0.007 &                       &      \\
                     &                      &  23.873 $\pm$  0.028 &                       &      \\
                     &                      &  24.000 $\pm$  0.006 &   24.06 $\pm$   0.20  & 0.16 \\
                     &                      &  24.117 $\pm$  0.010 &                       &      \\
                     &                      &  24.222 $\pm$  0.019 &                       &      \\
25.515 $\pm$  0.017  & 25.033 $\pm$  0.050  &  27.066 $\pm$  0.007 &                       &      \\
                     &                      &  27.223 $\pm$  0.024 &                       &      \\
                     &                      &  27.377 $\pm$  0.004 &   27.45 $\pm$   0.07  & 0.17 \\
                     &                      &  27.522 $\pm$  0.003 &                       &      \\
                     &                      &  27.674 $\pm$  0.007 &                       &      \\
                     &                      &  27.829 $\pm$  0.006 &                       &      \\
                     &                      &  28.009 $\pm$  0.004 &                       &      \\
29.186 $\pm$  0.004  & 28.593 $\pm$  0.079  &  30.605 $\pm$  0.016 &                       &      \\
                     &                      &  30.816 $\pm$  0.015 &                       &      \\
                     &                      &  31.041 $\pm$  0.003 &                       &      \\
                     &                      &  31.228 $\pm$  0.004 &   31.35 $\pm$   0.16  & 0.19 \\
                     &                      &  31.585 $\pm$  0.006 &                       &      \\
                     &                      &  31.810 $\pm$  0.021 &                       &      \\
                     &                      &  32.044 $\pm$  0.007 &                       &      \\
32.990 $\pm$  0.003  & 32.433 $\pm$  0.104  &  34.072 $\pm$  0.012 &                       &      \\
                     &                      &  34.323 $\pm$  0.006 &                       &      \\
                     &                      &  34.585 $\pm$  0.010 &                       &      \\
                     &                      &  34.833 $\pm$  0.008 &   34.90 $\pm$   0.09  & 0.29 \\
                     &                      &  35.055 $\pm$  0.004 &                       &      \\
                     &                      &  35.300 $\pm$  0.012 &                       &      \\
                     &                      &  35.576 $\pm$  0.012 &                       &      \\
36.633 $\pm$  0.005  & 36.079 $\pm$  0.154  &  38.355 $\pm$  0.030 &                       &      \\
                     &                      &  38.622 $\pm$  0.013 &   38.65 $\pm$   0.07  & 0.24 \\
                     &                      &  38.889 $\pm$  0.029 &                       &      \\
                     &                      &  39.199 $\pm$  0.019 &                       &      \\
40.298 $\pm$  0.038  &                      &  42.487 $\pm$  0.012 &                       &      \\
$^{*}$43.865 $\pm$  0.026 &                 &                      &                       &      \\
$^{*}$47.299 $\pm$  0.062 &                 &                      &                       &      \\
 \hline
 \multicolumn{5}{c}{RGB} \\
 \hline
59.328 $\pm$  0.332  & 58.427 $\pm$  0.337  &  69.971 $\pm$  0.011 &                       &      \\
66.404 $\pm$  0.036  & 65.395 $\pm$  0.080  &  70.206 $\pm$  0.007 &   70.25 $\pm$   0.16  & 0.04 \\
                     &                      &  70.422 $\pm$  0.007 &                       &      \\
73.678 $\pm$  0.007  & 72.733 $\pm$  0.033  &  77.023 $\pm$  0.008 &                       &      \\
                     &                      &  77.314 $\pm$  0.008 &   77.36 $\pm$   0.14  & 0.05 \\
                     &                      &  77.558 $\pm$  0.027 &                       &      \\
80.862 $\pm$  0.008  & 79.918 $\pm$  0.017  &  84.046 $\pm$  0.012 &                       &      \\
                     &                      &  84.472 $\pm$  0.012 &   84.54 $\pm$   0.29  & 0.06 \\
                     &                      &  84.762 $\pm$  0.016 &                       &      \\
88.374 $\pm$  0.204  &                      &                      &                       &      \\
\hline
\multicolumn{5}{l}{
  \parbox{\columnwidth}{
    \footnotesize Frequencies marked with $^{*}$ correspond to the two modes with an uncharacteristically small FWHM.}}
\end{tabular} \hfill~}
\vspace{2pt}

\end{table} 

\section{Model grid} \label{sec:model}

We generated our model grid with \mesa \citep[Modules for Experiments in Stellar Astrophysics;][version r24.03.1]{Paxton2011,Paxton2013,Paxton2015,Paxton2018,Paxton2019,Jermyn2023}.

Since the orbital separation of the two components of the binary system is large and interaction between them is negligible, we used single-star models and not the binary module implemented in \mesa.
We used the grey Eddington Temperature - opacity (T-$\tau$) relation \citep{Eddington1926} to describe the surface and allowed the opacity to vary to be consistent with the local temperature and pressure. We expanded the outer limit of the model from the photosphere to an optical depth of $10^{-3}$ to include the outer layers of the atmosphere to reduce the surface effect when calculating the oscillations (see \autoref{sec:surfEff_ball}). 
We included the pre-MS in our model calculations and roughly doubled the spatial and time resolution for models whose calculated \dnu was within $\pm20$ and $\pm30$ per cent of the observed value for the RGB and RC phase, respectively. We evolved our stellar models until the helium abundance was less than 0.001 per cent (by mass) of the total abundance in the core of the star.

Because of the advanced evolutionary state of our models, we considered all the reactions and isotopes included in \textit{`pp\_cno\_extras\_o18\_ne22.net'}. 
We did not consider rotation, atomic diffusion or gravitational settling. 

We used a Sobol' sequence \citep{sobol1967} of $2^{11}$ samples to distribute the following six grid parameters evenly over the parameter space. The different range of the varied parameters in our grid can be found in \autoref{tab:grid}. The choice of parameters and their ranges is explained below.

\begin{itemize}
\item{\textbf{Initial mass \Minit}}: We used the asteroseismic scaling relations \citep{Kjeldsen1995,Stello2009,Kallinger2010a}
\begin{equation}
\label{eq:scalingRel}
\frac{M_{*}}{M_{\odot}} \approx \left(\frac{\nu_{\max }}{\nu_{\max,\,\odot}}\right)^{3} \times\left(\frac{\Delta \nu}{\Delta \nu_{\odot} f_{\Delta\nu}}\right)^{-4} 
\times\left(\frac{T_{\mathrm{eff}}}{T_{\mathrm{eff,\,\odot}}}\right)^{3/2},
\end{equation}
to estimate the initial masses of the stars to constrain the mass range. We adopted the solar parameters from \citet{Huber2011} with $\nu_{\mathrm{max},\odot}=3090\pm30$\muHz, $\Delta\nu_{\odot}=135.1 \pm 0.1$\muHz, and a temperature of $T_{\mathrm{eff},\odot}= 5772.0 \pm 0.8 K$ \citep{Prsa2016}. 

We corrected \dnu by using the factor \fdnu \footnote{We note that $f_{\Delta\nu}$ is often defined as part of the nominator instead of the denominator in \autoref{eq:scalingRel}.} following the method introduced by \citet{LiY2023}, but use a different model grid including RC models (see \autorefA{sec:fdnu}). 

While \MH is the same for both stars, the same assumption cannot be made for \Teff. We do not have spectroscopic results for the RGB star, but using isochrones (see \autorefA{sec:isochrone}) to compare the two stars, we conclude that both stars have a similar temperature. We obtained a mass of $1.09\pm0.10$\,\Msun for the RC star and $1.01\pm0.12$\,\Msun for the RGB star.
We used these results to estimate the mass range of the grid, which we show in \autoref{tab:grid}.
\item{\textbf{Mass loss $\dot M$  }}; We implemented the Reimers law \citep{Reimers1975} to model stellar mass loss: 
\begin{equation} 
\dot M = 4 \times 10^{-13} \eta \frac{L R}{M},
\end{equation}
where $M$ is the mass, $L$ is the luminosity, and $R$ is the radius. The mass-loss rate is scaled by the free parameter $\eta$, which we varied in our grid. We note that the total mass loss varies according to how long the star spends on the tip of the RGB, and therefore depends strongly on the initial mass. We summarised the current state of research on mass-loss in red giants in the \autorefI and chose our range of $\eta$ accordingly.

\item{\textbf{Metallicity} $Z$}; 
We used the solar composition from \citet{Grevesse1998} as a reference to relate the metallicity $Z$ to \MH. We note that the metallicity was used as an input parameter of our grid, as well as one of the output parameters to constrain the models. Additionally, because both stars are formed from the same dust and gas cloud, we assume the stars share the same metallicity. 

\item{\textbf{He-abundance} $Y$}; The initial helium abundance is difficult to constrain for most stars, as there are generally no He absorption lines in the spectra. Nevertheless, it still plays a crucial part in the evolution of stars. It is common to infer this value by relating it to the metallicity and the primordial helium abundance through an enrichment law, but this is often considered to be an oversimplification \citep[e.g.][]{Nsamba2021}.
It is possible to more accurately estimate the He abundance by studying asteroseismic glitches caused by the He ionisation zones  \citep{Basu2004,Miglio2010,Verma2014,Verma2019,Verma2022}, but success is limited to ideal conditions.

\item{\textbf{Overshooting} \fov}; For our models, we chose the exponential overshooting scheme, which adds additional mixing at the border between convection and radiative zones in the core and envelope. In RC stars, the mixing at the border of the convective core remains an area of active research. The growth of the core during the helium burning phase depends strongly on the treatment of the convective boundary as discussed in the \autorefI. 
  
In future work, \mystar would also be an ideal candidate to explore different ways to treat the boundary mixing. However, different mixing schemes come with their own assumptions and uncertainties. For this paper, we decided to vary only the most commonly used exponential overshooting parameter and refrained from recalculating our grid with alternative overshoot descriptions. 

\item{\textbf{Mixing length} \aMLT}; In {\sc mesa}, the mixing length theory \citep{BohmVitenese1958} is used to describe convective mixing, giving the option of different implementations and extensions. For our models, we use the one described by \citep[][]{Kuhfuss1986}. The mixing length parameter \aMLT, which we vary in our grid, specifies how far a mass element can travel before it mixes with its surroundings. This value depends on the type of stars \citep[e.g.][]{Joyce2018b}. There are suggestions of a dependence on mass \citep{Trampedach2014} and also metallicity \citep{Viani2018,Tayar2017,Joyce2018}. While most find subsolar values for RGBs, some report they require supersolar values \citep[e.g.][]{Song2018,Li2018}. Results suggest a dependence on age and the evolutionary state, but there is no clear answer to how this dependence behaves. The choice of the mixing length parameter is still a large source of uncertainty in stellar modelling, and it is therefore good practice to allow for different values in model grids.
\end{itemize}

\begin{table}
\caption{Parameter ranges of the grid.}
\tabcolsep=6.5pt 
\hfill\begin{tabular}{c | l l} 
 \hline
 \hline
\multicolumn{1}{c}{parameter} &
\multicolumn{1}{c}{range}  \\

\hline
Mass [\Msun]          & 0.9 \ \ \ \ \     ---\ \ 1.3   \\
$\eta$                & 0.1 \ \ \  \ \    ---\ \ 0.3  \\
$Z$                   & 0.007 \           ---\ \ 0.022  \\
$Y$                   & 0.25 \ \ \        ---\ \ 0.3  \\
$f_{\mathrm{ov}}$     & 0 \ \ \ \ \ \ \ \ ---\ \ 0.02   \\
$\alpha_{\mathrm{MLT}}$& 1.6\ \ \ \ \ \   ---\ \ 2.3    \\
\hline

\end{tabular} \hfill~
\label{tab:grid}
\end{table}

We calculated the oscillation modes with the stellar oscillation code {\sc Gyre} \citep[][version 7.1]{Townsend2013}, as implemented within {\sc MESA}'s {\texttt{run\_star\_extras}} modules \citep[][]{Bellinger2022,Joyce++2024}. In addition to the $\ell=0$ modes, as mentioned in \autoref{sec:asteroseismology}, we did not calculate the mixed modes but instead estimated only the $\pi$ modes for $\ell=1$ and $2$, as implemented in the {\sc Gyre} code by \citet{Ong2020}. A few models had one extra $\pi$ mode for certain radial orders, but these extra modes were easy to identify and discarded based on their much higher inertias.

\section{Model fitting} \label{sec:statistics}
Both stars have similar masses and metallicities, and can be treated independently because of their large orbital separation. We therefore used a single grid of models, but we compared the primary only to the models of the red clump phase of evolution, and the secondary to the red giant branch phase. The observed and modelled individual oscillation frequencies were matched by taking the full set of observed modes and comparing them to a continuous set of modelled modes. We selected the modelled set that resulted in the smallest total weighted squared difference.  To reduce computational resources, we only calculated the oscillations when the large frequency separation of the model is within $\sim 20$ per cent of the observed value. For each of these models, we calculated the \X with
\begin{equation}
\chi_i^2=\frac{\left(x_{\mathrm{obs}}-x_{\mathrm{mod}}\right)^2}{\sigma_{\mathrm{obs}}^2}.
\end{equation}
The variable $x$ represents \MH for both stars, and additionally \Teff for the RC star and \dpi for the RGB star. The modelled value is indicated by $x{_\mathrm{mod}}$, and the observed value by $x_{\mathrm{obs}}$. 
In the following section, we describe how we calculated $\chi^2_\mathrm{freq}$ for the individual oscillation mode frequencies using two prescriptions to account for surface effects. But we note that no method has been tested for RC stars. Using two different ones may allow us to compare their performances.

\subsection{Surface correction - Ball \& Gizon method} \label{sec:surfEff_ball}

When modelling individual frequencies, we have to consider so-called surface effects. Observed modes differ from modelled modes by a function that depends on frequency, which is true for the Sun \citep{JCD1996} and for other stars \citet{Kjeldsen2008}. This deviation is due to invalid simplifications of the one-dimensional model of the surface and has to be corrected. Among the different approaches to correct these surface effects, the most commonly used method is described by \citet{Ball2014} using
\begin{equation} 
\delta \nu_{\mathrm{surf}}=\left(a_{-1}\left(\nu / \nu_{\mathrm{ac}}\right)^{-1}+a_3\left(\nu / \nu_{\mathrm{ac}}\right)^3\right) / \mathcal{I}.
\end{equation}
The surface term \dnusurf describes the difference between the modelled and observed mode frequencies at a given frequency $\nu$. The quantities $\mathcal{I}$ and $\nu_{\mathrm{ac}}$ are the mode inertia and the acoustic cutoff frequency, while $a_{-1}$ and $a_{3}$ are two coefficients, estimated by fitting the surface-corrected modelled modes to the observed radial modes and thus differ for each model. We calculated the \X with 
\begin{equation}
\chi_{\mathrm{freq},B14}^2=\frac{1}{N_{\mathrm{modes}}-2} \sum_{i}^{N_{\mathrm{modes}}}\left(\frac{\nu_i ^\mathrm{mod}+\delta v_{\mathrm{surf},i}-v_i^\mathrm{obs}}{\sigma^\mathrm{obs}_i}\right)^2,
\end{equation}
with $\sigma_{i}^\mathrm{obs}$ being the uncertainties of the observed modes, and \X normalized, by $N_{\mathrm{modes}} -2$, the number of modes minus the number of fitted coefficients.
The \citet{Ball2014} method has been successfully applied in numerous studies, often outperforming other approaches that depended on parameterisation \citep{Schmitt2015,Ball2017,Basu2018,Compton2018,Jorgensen2020}. However, it often needs additional constraints to avoid unreasonably large correction terms. 

Modes with a low radial order ($n$) are less affected by the surface effects due to their longer wavelengths, and the surface term is expected to grow with increasing radial order. Following the results observed for the Sun and other stars, the correction term is also assumed to be positive. In a large-scale study, \citet{LiY2023} found a dependence on the evolutionary state of the stars, and their results can be used to further constrain the parameter-dependent correction. RC stars were excluded from these studies, and the surface term for these stars is thus untested and unconstrained. We note that the surface term is also dependent on the description of the model atmosphere.

\subsection{Surface correction - Roxburgh method}
\label{sec:surfEff_rox}
An alternative method to deal with the poorly modelled surface, without the need for fitting models to the observations, was described by \citet{Roxburgh2016}, and focuses on the phase shift, \epsp. Similar to the method of comparing ratios of different frequency separations, as described by \citet{Roxburgh2003}, this approach does not "correct" the modelled frequencies but instead relies on properties unaffected by the star's surface. The \citet{Roxburgh2016} method is therefore independent of the model atmosphere and its systematic errors. It is not commonly used---nor as well tested---as the method from \citet{Ball2014}, especially for red giants, but was successfully applied by \citet{Ong2021} and \citet{Campante2023}.
The method uses the fact that oscillation modes with different angular degrees probe different stellar depths, but all interact similarly with the surface layers. 
Consequently, the surface contribution to \epspp is independent of angular degree $\ell$ and depends only on the frequency. 
Assuming the model accurately represents the stellar interior, subtracting the observed phase shift from the corresponding modelled phase shift cancels the $\ell$-dependent interior contribution, leaving only the $\ell$-independent surface contribution.

We interpolated the modelled modes to the frequencies of the observations and subtracted them from the observations:
\begin{equation}\label{eq:rox_epsdiff}
\mathcal{E}(\nu_{i}^\mathrm{obs})=\epsilon_{\mathrm{p,i}}^{\prime\,\mathrm{mod}}\left(\nu_{i}^{\mathrm{obs}}\right)-\epsilon_{\mathrm{p},i}^{\prime\,\mathrm{obs}}\left(\nu_{i}^{\mathrm{obs}}\right).
\end{equation}
We fitted the resulting differences, $\mathcal{E}(\nu_{i}^\mathrm{obs})$ with Chebychev polynomials to obtain an $\ell$-independent function, $\mathcal{F}\left(v_{i}\right)$. We estimated \X, following \citet{Roxburgh2016}, with 
\begin{equation}
\chi_{\mathrm{freq},R16}^2=\frac{1}{N_{\mathrm{modes}}-M} \sum_i^{N_{\mathrm{modes}}}\left(\frac{\mathcal{E}\left(\nu_{i}^\mathrm{obs}\right)-\mathcal{F}\left(\nu_{i}^\mathrm{obs}\right)}{\sigma_{i}^\mathrm{obs}/\Delta \nu} \right)^2,
\end{equation}
where $M$ is the number of parameters in $\mathcal{F}$, which we selected to be one less than the number of $\ell=0$ modes.

\subsection{Binary constraints} \label{sec:binaryChi}

We calculated the total \X for the RC and RGB star:
\begin{equation}
\chi^2_\mathrm{RC}=\chi^2_\mathrm{\mathrm{freq}}+\chi^2_\mathrm{Teff}+\chi^2_\mathrm{Z},
\end{equation}
\begin{equation}
\chi^2_\mathrm{RGB}=\chi^2_\mathrm{\mathrm{freq}}+\chi^2_\mathrm{Z}+\chi^2_\mathrm{\Delta\Pi}.
\end{equation}
Next, we define our constraints that we obtained because the stars are part of the same binary system. We define an additional \X that compares the ages and compositions of both stars. 
\begin{equation}
\chi^2_\mathrm{binary}=\frac{\left(t_\mathrm{RC}-t_\mathrm{RGB}\right)^2}{s_t^2} + \frac{\left(Z_{\mathrm{RC}}-Z_{\mathrm{RGB}}\right)^2}{s_{_{\mathrm{\mathrm{Z}}}}^2} + \frac{\left(Y_{\mathrm{RC}}-Y_{\mathrm{RGB}}\right)^2}{s_{\mathrm{Y}}^2}
\end{equation}
We did not interpolate the grid, and therefore, discrete timesteps introduce an age uncertainty when comparing the age of both stars. We defined the uncertainty $s_t^2$, which depends on the largest timestep in the tracks of the corresponding evolutionary state of both models, multiplied by the arbitrary value 10. We choose to downweight the contribution of the age difference because of the high uncertainties associated with the stellar age obtained from models, and to not impose an unrealistic strong constraint. The uncertainties of the initial metallicity, $s_{Z}$, and He abundances, $s_{Y}$, are set by the resolution of the grid,
\begin{equation}
s_{Z}=\frac{Z_{\max }-Z_{\min }}{N_{\text {track }}^{1/{N_{\text {param }}}}},
\end{equation}
with the number of tracks $N_{\text {track}}=2^{11}$ and the number of varied initial parameters in our grid $N_{\text {param}}=6$ (Chiu in prep.; \citealt{Chiu2025}). This results in a $s_{Z}$ of $0.006$ and a $s_{Y}$ of $0.020$. For $s_{t}$ the value depends on the model, but it is typically of an order of $\sim10$\,Myr.  

We use the values of $\chi^2_\mathrm{RC}$ and $\chi^2_\mathrm{RGB}$ to obtain the model parameters, representing the solution if we assume a single star system. For our solutions assuming a binary system, we added $\chi^2_\mathrm{binary}$ to every possible combination of all $\chi^2_\mathrm{RC}$ and $\chi^2_\mathrm{RGB}$ for all models. However, the inherently larger values of $\chi^2_\mathrm{RC}$ would dominate the combined \X, forcing the results for the RGB models to align with those of the RC models.
To avoid this and to account for the larger systematic uncertainties in RC models, we rescaled the RC and RGB \X values by dividing them all by the corresponding best-fitting single-star model.
The end result is that both stars contribute approximately equally to the combined \X. The exact choice of weighting of both stars and of the age and composition differences has little impact on our final conclusions. The results obtained using the binary constraints mainly serve as a quality check of our single-star model results rather than to estimate accurate parameters.
Furthermore, to prevent duplicate models in our final analysis, we selected for each individual RGB (or RC) evolutionary track the best-fitting model and then searched for the companion star model that resulted in the lowest  $\chi^2_\mathrm{RGB}+\chi^2_\mathrm{RC}+\chi^2_\mathrm{binary}$. The companion star model was chosen from a selection of only the best-fitting models of each track. This last constraint was necessary to avoid pairs that had the same age and composition, but where both models individually were bad representations of the stars themselves. 
This way, we kept the original number of models and corresponding \X values for each star for our results when treating the components as single stars or as part of a binary.

From \X, we calculated the likelihood,
\begin{equation}
\mathcal{L}(x)=\exp \left(-\frac{1}{2} \chi^2\right),
\end{equation}
and the probability distribution, which we approximated by a weighted histogram, with the weights being the likelihood of the model.
Finally, we normalized this distribution to ensure that the sum of the bin weights equals one.

\section{Results and discussion} \label{sec:results}
\subsection{Ball \& Gizon method - constraining the surface term}
The \citet{Ball2014} surface correction method often requires constraints to prevent physically unrealistic solutions, because unconstrained models frequently result in a low \X but with an unexpectedly large, negative, or a frequency-decreasing, \dnusurf. To prevent this, we added constraints that disregard models for which \dnusurf or its derivative have the wrong sign.
Furthermore, studies of the Sun show that \dnusurf becomes negligible for low-order radial modes \citep{JCD1996}. We therefore excluded models where the frequency difference between the lowest observed radial mode and the surface-corrected modelled mode exceeds $0.1$\dnu.

\begin{figure*}
  \includegraphics[width=0.5\linewidth,trim={0 0.2cm 0 1.05cm},clip]{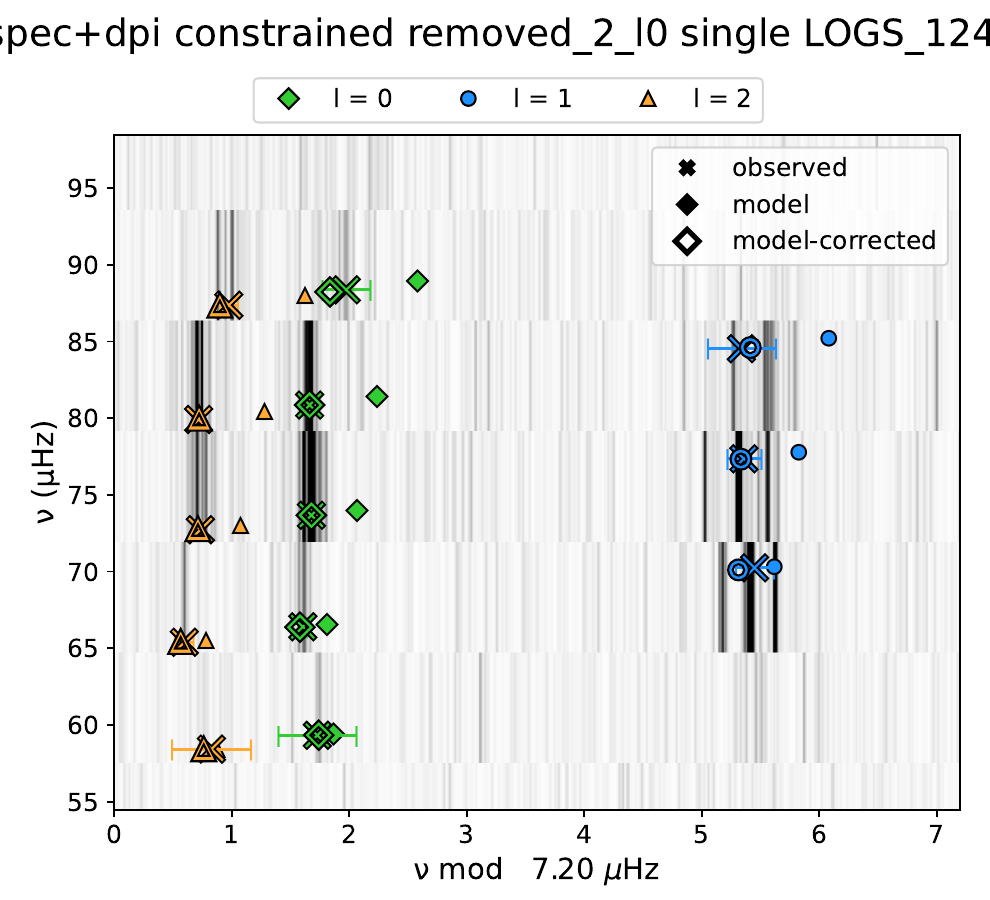}\hfill
  \includegraphics[width=0.5\linewidth,trim={0 0.2cm 0 1.15cm},clip]{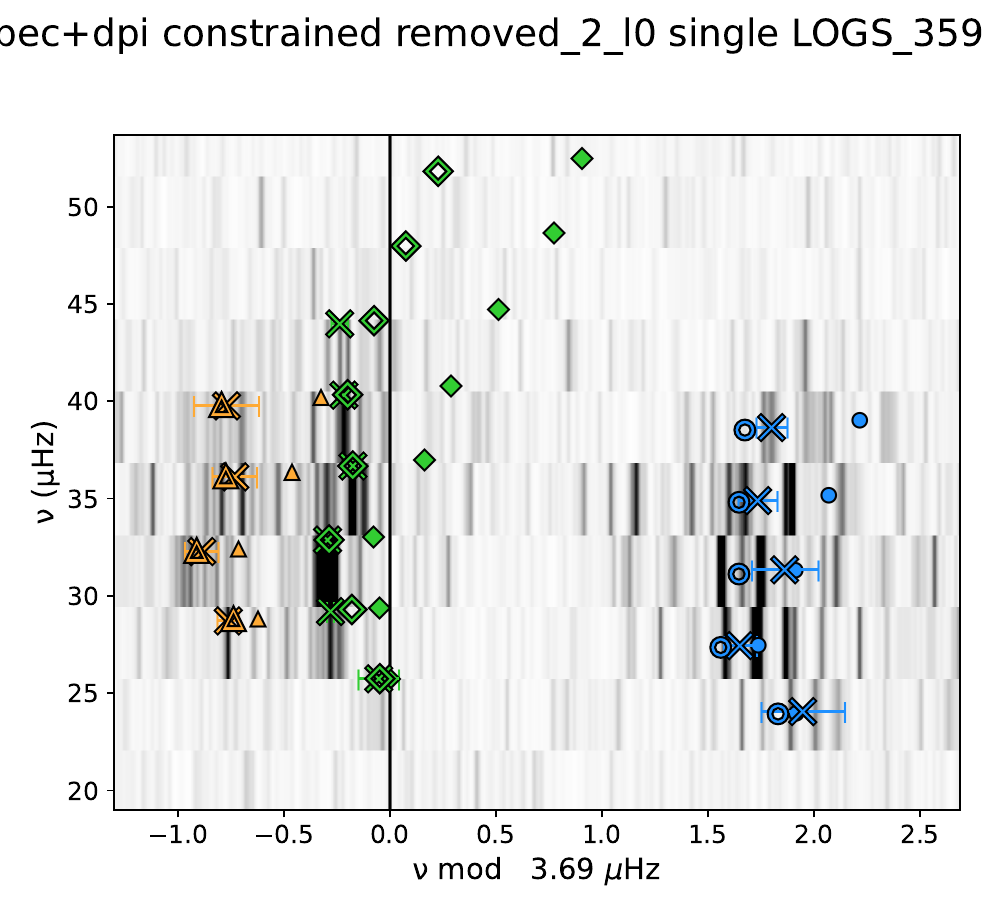}
  \caption{\Echelle diagrams of the red giant branch (left panel) and red clump (right panel) components of \mystar, comparing the observed mode frequencies (crosses) and model frequencies before and after correcting for the surface effects (filled and empty symbols, respectively).}
  \label{fig:RC_badfit}
\end{figure*}

\begin{figure}
  \includegraphics[width=\linewidth,trim={0 0.9cm 0 1.05cm},clip]{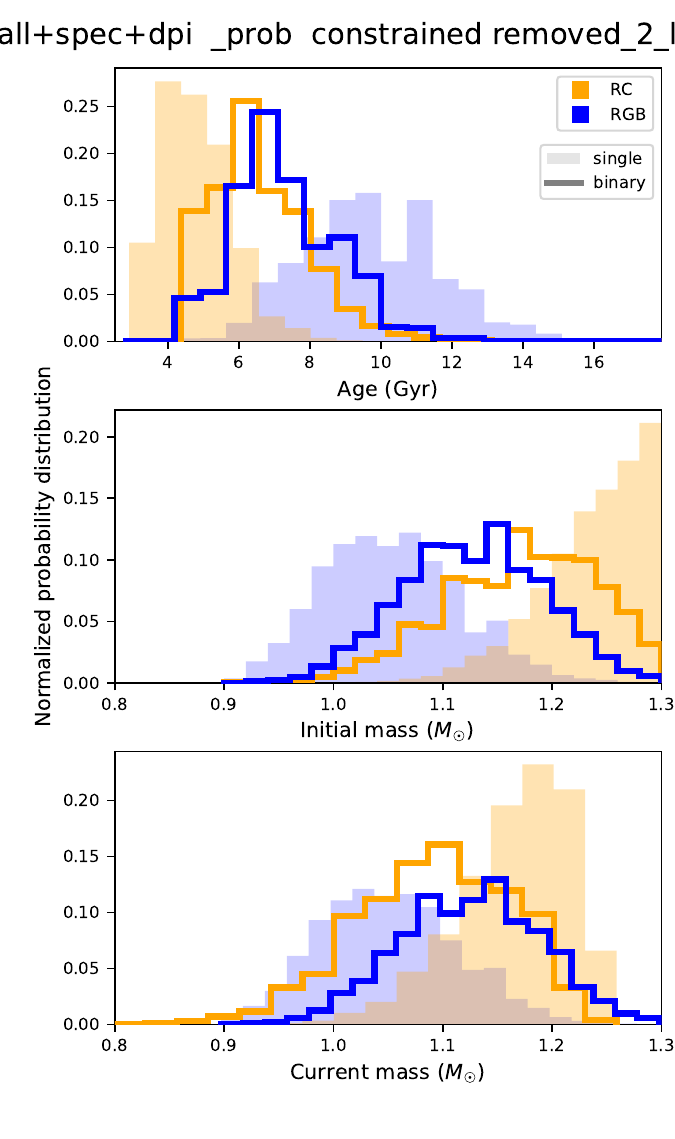}
  \caption{Normalised probability distribution of the age (top panel), initial mass (middle panel) and current mass (bottom panel) of both components of \mystar. The surface correction term was constrained as described in the text. The colour blue represents the red giant branch star, and the colour orange the red clump star. The shaded region corresponds to the result when both stars are treated independently from each other, and the solid line when additional constraints---common age and initial composition---are added. }
  \label{fig:probdis_bad}
\end{figure}

We show the results for the RGB star, treated as a single star, in \autoref{fig:RC_badfit} and the probability distribution in \autoref{fig:probdis_bad} as the shaded blue area. We used a kernel density estimation (KDE) to obtain the stellar parameters with the highest density and show them in \autoref{tab:results}.
We report the $16^{\mathrm{th}}$ and $84\mathrm{th}$ percentile as the uncertainties.
Finally, we note that the probability distribution of the helium abundance and the overshooting parameter was mostly flat, and we were unable to constrain it, which can also be seen in the corner plots in \autorefA{sec:cornerplots}.

We found RC models with reasonably good results for the surface correction using these additional constraints, but even the best-fitting RC models had very high \XRC $>130$ (whereas the best \XRGB $ < 0.1$). This is mainly due to a poor fit of the modelled to the observed individual modes, especially the two highest-frequency radial modes. The two corresponding Lorentzians in the PSD also seem to be much narrower than what is expected from the lifetime of these modes, especially when compared with the other radial modes (see \autoref{fig:streched}). On the other hand, they do follow the expected glitch structure of red clump stars as shown in \citep[e.g.][]{Vrard2015}. Ignoring these two modes improved the fit (see \autoref{fig:RC_badfit}), but ultimately gave a still relatively high \XRC $>5$. This is shown by a comparison of the normalized probability distribution of the age of the RC to the RGB star, which we show in \autoref{fig:probdis_bad}. 

While the results obtained by treating the system as a binary---constraining age and composition---appear in agreement with our expectations, they differ significantly from the results when treating the stars separately. Ideally, in a robust fit the "single"-star and "binary" results should overlap, yielding comparable ages even without explicitly enforcing this constraint. Although the RC fit alone looks acceptable (see \autoref{fig:RC_badfit}), its resulting age is incompatible with that of the RGB companion, even after accounting for large uncertainties. Such a massive discrepancy cannot be explained by typical systematic modelling errors (e.g. uncertain mixing processes), since we modelled both stars with the same code and input physics. Any systematic effects would influence both models similarly, especially during their main-sequence evolution, where they spend by far most of their lifetime. Further, the RC results imply that we underestimated its mass, which contradicts both the scaling relations and the expected age of the system. 

To rule out errors arising from using the $\pi$ modes, we repeated our calculations using different combinations of oscillation modes and additionally checked alternative definitions of \X, for example, ignoring $\chi^2_\mathrm{Z}$ and $\chi^2_\mathrm{Teff}$. The results were consistent within the uncertainties and suggest that our RC models are not able to accurately represent the star, regardless of the approach used. 

\subsection{A possible \texorpdfstring{$\epsilon_\mathrm{p}$}{epsilon} offset} 
\label{sec:epsilonOffset}
\begin{figure}
  \includegraphics[width=\linewidth,trim={0 0.2cm 0 1.05cm},clip]{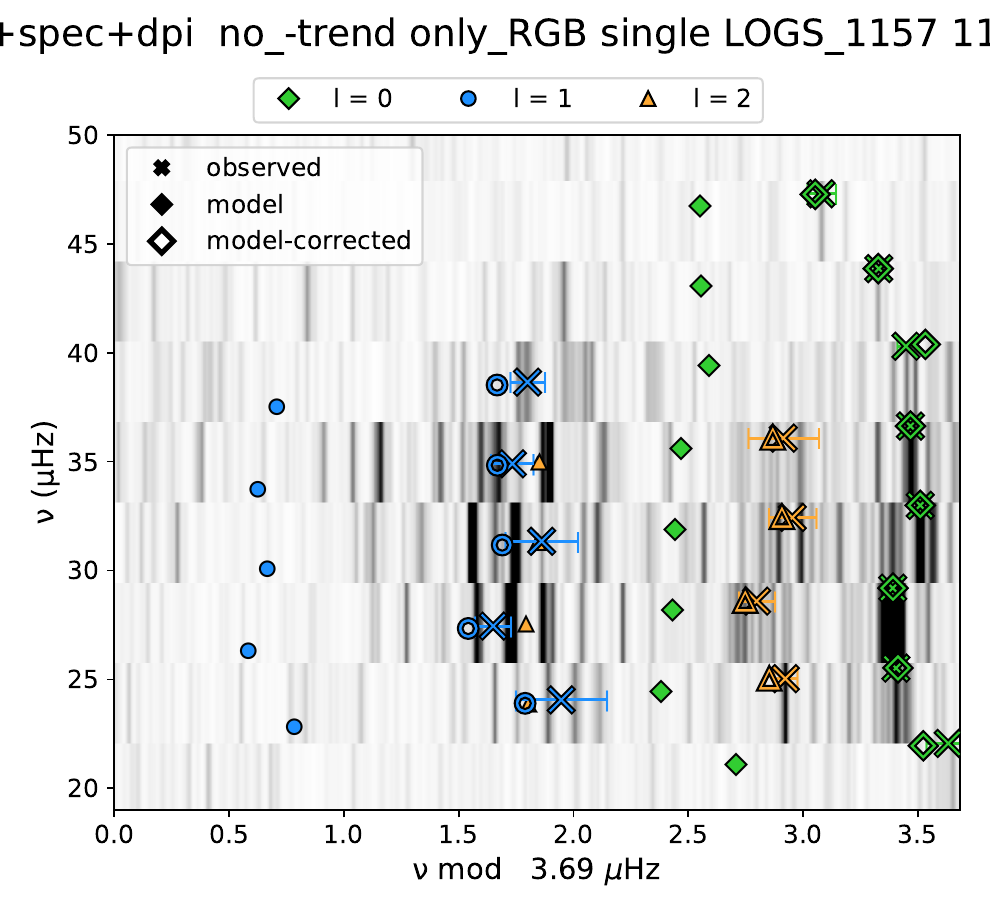}
  \caption{\Echelle of the red clump component. Same as \autoref{fig:RC_badfit} but the surface term was constrained less strictly (see \autoref{sec:epsilonOffset}).}
  \label{fig:RC_good}
\end{figure}

To investigate the disagreement between observations and models for the RC star, we recalculated the probability distributions with a different set of constraints for the surface correction, \dnusurf. We found promising results when we retained the constraint to remove negative trends---smaller \dnusurf for higher frequencies than for lower frequencies---but allowed models with a positive \dnusurf, as we show in \autoref{fig:RC_good}. We refer to these models as having an `epsilon offset'.

\begin{figure}
  \includegraphics[width=\linewidth,trim={0 0 0 1.05cm},clip]{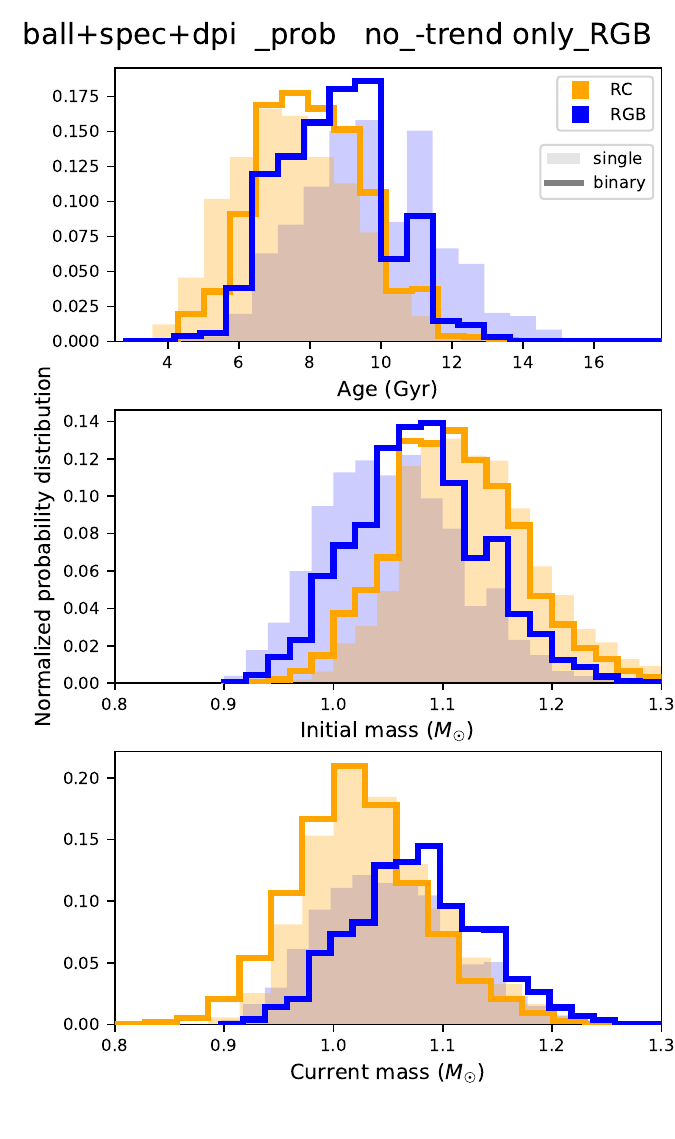}
  \caption{Normalised probability distribution. Same as \autoref{fig:probdis_bad}, but when the surface term is less strictly constrained for the red clump component (see \autoref{sec:epsilonOffset}). }
  \label{fig:probdis_good}
\end{figure}

Overall, the model in \autoref{fig:RC_good} reproduces the observed glitch pattern well across all angular degrees and with the inclusion of the epsilon offset, the individual frequencies fit well within the typical uncertainties for modelling RGs. Although the dipole and quadrupole modes seem to require a slightly different offset. Moreover, the age of these models agrees much better with that of the RGB companion (see \autoref{fig:probdis_good}), and there is no need anymore to ignore the two higher frequency radial modes. We decided to include them and regard them as real because they fitted well to the model, even when we didn't consider them in the fit. While these results look promising, we are cautious about their interpretation. Currently, we have not yet determined the origin or implications of this \epsp offset. More research needs to be done on similar stars to confirm that we can even replicate this \epsp offset and to explore its potential for improving RC modelling. 

For completeness, we also tested the RGB star without constraining the surface effects. While for some good fitting models \dnusurf is slightly positive and the overall uncertainties predictably larger, the final stellar parameters remained consistent within their uncertainties, regardless of the treatment. We looked at a few of the models with a positive \dnusurf in more detail. Each of them had a nearby model in the same evolutionary track (with an age difference of $\Delta t \ll 0.01$\,Gyr) with a slightly higher \X but a reasonable \dnu. This hints that the glitch structure of the oscillation modes is more indicative of the star's structure and parameters than the frequencies themselves or the method to constrain surface corrections. This is also discussed by Kallinger (submitted). In other words, for basically every evolutionary track, it is possible to find an RGB model with a more-or-less reasonable surface correction, regardless of how bad the model is (as we can also see for the RC star). A model's goodness of fit is more strongly determined by its ability to reproduce the glitch structure relating the frequencies together, rather than by the frequencies themselves.

\begin{table*}
\caption{The peaks of the KDE of the probability distributions of different model parameters. }
\label{tab:results}
\begin{tabular}{ccrrrrrrrr}
\hline
\hline
& & \multicolumn{1}{c}{age} & 
    \multicolumn{1}{c}{initial mass} & 
    \multicolumn{1}{c}{current mass} & 
    \multicolumn{1}{c}{\Teff} & 
    \multicolumn{1}{c}{Z} & 
    \multicolumn{1}{c}{Y} & 
    \multicolumn{1}{c}{$\alpha_{\mathrm{MLT}}$} & 
    \multicolumn{1}{c}{\dpi} \\
& & \multicolumn{1}{c}{[Gyr]} & 
    \multicolumn{1}{c}{[\Msun]} & 
    \multicolumn{1}{c}{[\Msun]} & 
    \multicolumn{1}{c}{[K]} & 
    \multicolumn{1}{c}{-} & 
    \multicolumn{1}{c}{-} & 
    \multicolumn{1}{c}{-} & 
    \multicolumn{1}{c}{[s]} \\
\hline
\multirow{3}{*}{single} 
& RGB & 
$   9.1 $ {\raisebox{0.5ex}{\tiny$\substack{ + 1.4 \\ - 2.6  } $ } } & 
$  1.041 $ {\raisebox{0.5ex}{\tiny$\substack{ +0.053 \\ -0.079  } $ } } & 
$  1.039 $ {\raisebox{0.5ex}{\tiny$\substack{ +0.053 \\ -0.079  } $ } } & 
$  4619 $ {\raisebox{0.5ex}{\tiny$\substack{ +  67 \\ - 134  } $ } } & 
$  0.0180 $ {\raisebox{0.5ex}{\tiny$\substack{ +0.0023 \\ -0.0021  } $ } } & 
$  0.288 $ {\raisebox{0.5ex}{\tiny$\substack{ +0.027 \\ -0.004  } $ } } & 
$  1.83 $ {\raisebox{0.5ex}{\tiny$\substack{ +0.12 \\ -0.19  } $ } } & 
$  72.62 $ {\raisebox{0.5ex}{\tiny$\substack{ +0.23 \\ -0.91  } $ } } 
\\
& RC & 
$  4.39 $ {\raisebox{0.5ex}{\tiny$\substack{ +0.71 \\ -1.33  } $ } } & 
$  1.273 $ {\raisebox{0.5ex}{\tiny$\substack{ +0.061 \\ -0.026  } $ } } & 
$  1.190 $ {\raisebox{0.5ex}{\tiny$\substack{ +0.065 \\ -0.040  } $ } } & 
$  4850 $ {\raisebox{0.5ex}{\tiny$\substack{ + 160 \\ - 130  } $ } } & 
$  0.0182 $ {\raisebox{0.5ex}{\tiny$\substack{ +0.0072 \\ -0.0021  } $ } } & 
$  0.292 $ {\raisebox{0.5ex}{\tiny$\substack{ +0.020 \\ -0.006  } $ } } & 
$  1.97 $ {\raisebox{0.5ex}{\tiny$\substack{ +0.14 \\ -0.20  } $ } } & 
$   296 $ {\raisebox{0.5ex}{\tiny$\substack{ +  17 \\ -  15  } $ } } 
\\
& RC (\epsp offset) &
$   7.1 $ {\raisebox{0.5ex}{\tiny$\substack{ + 1.4 \\ - 2.1  } $ } } & 
$  1.113 $ {\raisebox{0.5ex}{\tiny$\substack{ +0.047 \\ -0.075  } $ } } & 
$  1.020 $ {\raisebox{0.5ex}{\tiny$\substack{ +0.041 \\ -0.082  } $ } } & 
$  4790 $ {\raisebox{0.5ex}{\tiny$\substack{ + 130 \\ - 120  } $ } } & 
$  0.0176 $ {\raisebox{0.5ex}{\tiny$\substack{ +0.0057 \\ -0.0021  } $ } } & 
$  0.2619 $ {\raisebox{0.5ex}{\tiny$\substack{ +0.0037 \\ -0.0268  } $ } } & 
$  1.97 $ {\raisebox{0.5ex}{\tiny$\substack{ +0.17 \\ -0.21  } $ } } & 
$   285 $ {\raisebox{0.5ex}{\tiny$\substack{ +  26 \\ -  12  } $ } } 
\\
\hline
\multirow{2}{*}{binary} 
& RGB    &
$   8.7 $ {\raisebox{0.5ex}{\tiny$\substack{ + 1.6 \\ - 1.7  } $ } } & 
$  1.076 $ {\raisebox{0.5ex}{\tiny$\substack{ +0.062 \\ -0.066  } $ } } & 
$  1.074 $ {\raisebox{0.5ex}{\tiny$\substack{ +0.062 \\ -0.066  } $ } } & 
$  4610 $ {\raisebox{0.5ex}{\tiny$\substack{ +  48 \\ - 144  } $ } } & 
$  0.0168 $ {\raisebox{0.5ex}{\tiny$\substack{ +0.0021 \\ -0.0032  } $ } } & 
$  0.276 $ {\raisebox{0.5ex}{\tiny$\substack{ +0.015 \\ -0.014  } $ } } & 
$  1.81 $ {\raisebox{0.5ex}{\tiny$\substack{ +0.12 \\ -0.18  } $ } } & 
$  72.62 $ {\raisebox{0.5ex}{\tiny$\substack{ +0.18 \\ -0.95  } $ } } 
\\
& RC (\epsp offset) &  
$   7.8 $ {\raisebox{0.5ex}{\tiny$\substack{ + 1.4 \\ - 1.8  } $ } } & 
$  1.105 $ {\raisebox{0.5ex}{\tiny$\substack{ +0.054 \\ -0.067  } $ } } & 
$  1.017 $ {\raisebox{0.5ex}{\tiny$\substack{ +0.055 \\ -0.066  } $ } } & 
$  4790 $ {\raisebox{0.5ex}{\tiny$\substack{ + 160 \\ - 100  } $ } } & 
$  0.0185 $ {\raisebox{0.5ex}{\tiny$\substack{ +0.0052 \\ -0.0017  } $ } } & 
$  0.2632 $ {\raisebox{0.5ex}{\tiny$\substack{ +0.0046 \\ -0.0256  } $ } } & 
$  2.02 $ {\raisebox{0.5ex}{\tiny$\substack{ +0.21 \\ -0.17  } $ } } & 
$   286 $ {\raisebox{0.5ex}{\tiny$\substack{ +  25 \\ -  12  } $ } } 
\\
\hline
\end{tabular}
\end{table*}

\subsection{The Roxburgh method}
We also explored the alternative approach of treating surface effects suggested by \citet{Roxburgh2016} (see \autoref{sec:surfEff_rox}). We found two distinct solutions for the RGB star. One gives a mass of 1.09\,\Msun, which is in good agreement with the result from the \citet{Ball2014} method. 
The second solution had a lower mass of 0.93\,\Msun and we found no clear hint in other stellar parameters that would distinguish between these two solutions.

\begin{figure}
  \includegraphics[width=\linewidth,trim={0 0 0 0},clip]{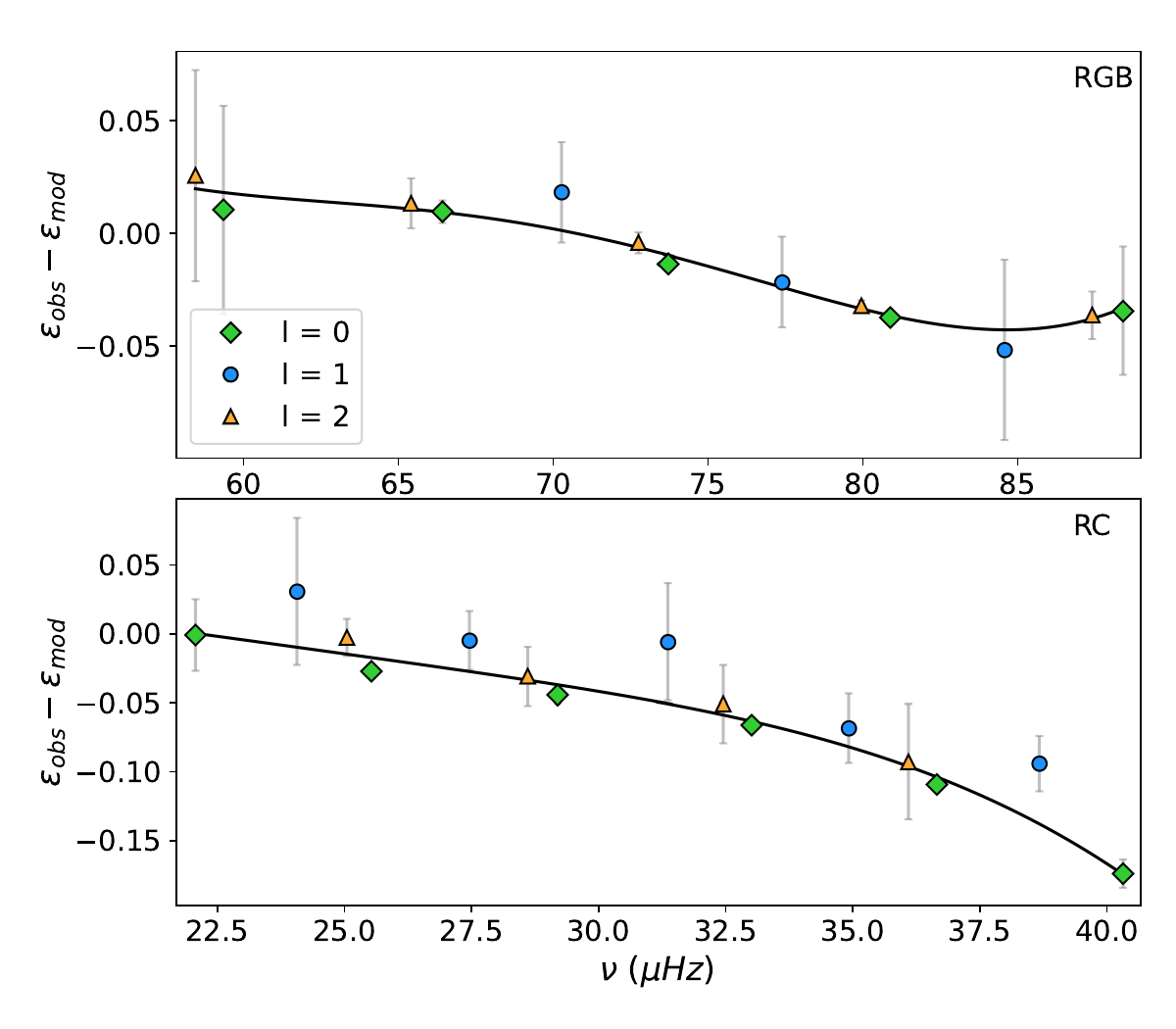}
  \caption{The difference between the phase shift of the observations and the model interpolated to the frequencies of the observed mode frequencies (see \autoref{eq:rox_epsdiff}). The black line corresponds to the Chebychev polynomials fit, $\mathcal{F}$. If the model matches the observations, the fit is independent of $\ell$.}
  \label{fig:rox}
\end{figure}

For the RC star, we were unable to find a good fit. The best-fitting models were spread widely over the whole parameter range. The probability distributions did not show a robust peak and were highly sensitive to slight changes in the weighting of the oscillation modes or other parameters. This is in agreement with our previous results that standard RC models are insufficient to describe real stars. In our specific case, the \citet{Roxburgh2016} method did not provide any further valuable insights into the properties of RC stars or their models.
However, we highlight its potential because it does not rely on parameterisation to fit models to observations. With the \citet[]{Ball2014} method, it is possible to find misleading solutions even when the model does not represent the observations well, as seen in \autoref{fig:RC_badfit}. Our binary system provided a great opportunity to identify these unreliable solutions, which is not available for other studies. Combining the \citet[]{Roxburgh2016} method with the \citet[]{Ball2014} method may be a promising way to provide a quality check in such cases. 
In \autoref{fig:rox} we show the results from \autoref{eq:rox_epsdiff} for the single-star results with the constrained surface correction, together with the corresponding Chebychev polynomial fit (see \autoref{sec:surfEff_rox}). For the RC star, we can see a $\ell$-dependent deviation from the fit, which implies a discrepancy between the inner structure of the model and the star. We note that while the $\ell=0$ and $\ell=2$ modes seem both to align with the fit, there is still a systematic offset, and the frequency dependence appears $\ell$ dependent. This illustrates how we can identify models not agreeing with observations, even when the modes seem to match after applying the surface correction described by \citet{Ball2014}.

\subsection{A comparison of the modelled and observed \texorpdfstring{\dpi}{Delta Pi} }
We note that the observed \dpi$=236.9$\,s of the RC star is significantly lower than the \dpi$\sim290$\,s predicted by our models. This mismatch in \dpi is in the opposite direction to previous studies \citep[e.g.][]{Constantino2015,Noll2025}, which found that expanding the convective core was necessary to increase the modelled \dpi to match them to observations. We instead find that our observed \dpi of the RC in \mystar is smaller than that of the models. However, those previous studies compared the distribution of RC stars and not individual stars. They showed that the distribution of \kepler RC stars peaks at \dpi $> 300$\,s for stars of about $1$\,\Msun, whereas standard overshooting models tend to produce values below 300\,s (see e.g. Fig.1 in \citealt{Constantino2015}). The RC in our study has an observed \dpi on the lower edge of the observed \kepler sample \citep[e.g][]{Vrard2016} and therefore lies even below the average \dpi predicted by standard overshooting models. This could also imply that enlarging the convective core in models may not be the full solution across the complete range of RC stars.
\section{Conclusions} \label{conclusion}
In this work, we analysed the asteroseismic binary \mystar, which contains a red giant branch (RGB) and a red clump (RC) star. Because both components of the binary system \mystar share the same age and initial composition, we can use the modelling results obtained for the RGB star to estimate the accuracy of the RC model. Whereas most studies focus on the g-mode period spacing to constrain RC models, we aimed for the opposite approach and focused on using the p modes to test RC models.     

Spectroscopy confirmed the system as a binary and not a chance alignment of two oscillating stars. We estimated \Teff and \MH of the primary with limited success due to the entanglement of the spectra of both stars.
We obtained the global seismic parameters and oscillation modes of both stars and used the stretched period and frequency \echelle diagrams to decouple the dipole mixed modes to obtain the pure p modes (so-called $\pi$ modes). This allows us to analyse the star independently of the period spacing but without completely disregarding the dipole modes.
We tested different approaches to assess our models and explored the \citet{Ball2014} and \citet{Roxburgh2016} methods to account for the surface effects. We found that conventional modelling could not reproduce the RC observations while remaining consistent with the RGB age. However, by using the \citet{Ball2014} method in an unconventional way, we discovered a possible solution: a large, possibly $\ell$-dependent, offset of the p-mode phase shift (\epsp) that we identified by allowing the surface correction to be positive. This does not necessarily mean the surface is modelled inaccurately, but instead may indicate a problem with the treatment of the convective core boundary and its effects on the total structure of the star. Further investigations are planned. 

Our results demonstrate that exploring g modes and the period spacing (\dpi) are not the only methods to investigate and possibly improve RC-models. Not only the core, which is sensitive to g modes, but also the star's outer layers, sensitive to p modes, are affected and hold valuable information. Not only \dpi, but also \epsp in RC-models disagree with observations. We show that RC modelling could lead to results with high systematic errors and should only be interpreted critically and with scepticism. 

We intend to expand our research to a larger sample of oscillating RC stars to consider the possibility of this system being an outlier and to determine whether the `epsilon offset' is an anomaly specific to this system. In particular, RC stars in stellar clusters provide similar strong constraints.
In this work, we focused on the common method of describing the convective core boundary using exponential overshooting and we plan to explore different options in a future project. We aim to investigate the potential of combining  \dpi and \epsp as diagnostic tools to test the treatment of the convective core boundary in RC stars.

\section{Acknowledgements} 

This paper includes data collected by the \kepler mission and obtained from the MAST data archive at the Space Telescope Science Institute (STScI). Funding for the \kepler mission is provided by the NASA Science Mission Directorate. STScI is operated by the Association of Universities for Research in Astronomy, Inc., under NASA contract NAS 5–26555.
This paper is based on observations obtained with the HERMES spectrograph, which is supported by the Research Foundation - Flanders (FWO), Belgium, the Research Council of KU Leuven, Belgium, the Fonds National de la Recherche Scientifique (F.R.S.-FNRS), Belgium, the Royal Observatory of Belgium, the Observatoire de Genève, Switzerland and the Thüringer Landessternwarte Tautenburg, Germany.
This research is based in part on data collected at the Subaru Telescope, which is operated by the National Astronomical Observatory of Japan. 
This research was undertaken with the assistance of resources from the National Computational Infrastructure (NCI Australia), an NCRIS enabled capability supported by the Australian Government.

LSS, TRB, and CLC acknowledge support from the Australian Research Council through Laureate Fellowship FL220100117. 
PGB acknowledges support by the Spanish Ministry of Science and Innovation with the \textit{Ram{\'o}n\,y\,Cajal} fellowship number RYC-2021-033137-I and the number MRR4032204. PGB, DHG and RAG acknowledge support from the Spanish Ministry of Science and Innovation with the grant no. PID2023-146453NB-100 (\textit{PLAtoSOnG}). MGP acknowledges support from the Professor Harry Messel Research Fellowship in Physics Endowment, at the University of Sydney. DHG acknowledges the support of a fellowship from ”la Caixa” Foundation (ID 100010434). The fellowship code is LCF/BQ/DI23/11990068.
DH acknowledges support from the National Aeronautics and Space Administration (80NSSC22K0781) and the Australian Research Council (FT200100871).
RAG acknowledges the support from the PLATO Centre National D'{\'{E}}tudes Spatiales grant. SM acknowledges support from the Spanish Ministry of Science and Innovation (MICINN) with the grant No. PID2023-149439NB-C41.

\textit{Software:} 
\texttt{numpy} \citep{numpy2020},  
\texttt{matplotlib} \citep{Matplotlib2007},  
\texttt{scipy} \citep{SciPy2020},
This research made use of \texttt{Astropy} \citep{astropy2013, astropy2018}, a community-developed core Python package for Astronomy. 

This work has utilised the stellar evolutionary code package, Modules for Experiments in Stellar Astrophysics
\citep[MESA][]{Paxton2011, Paxton2013, Paxton2015, Paxton2018, Paxton2019, Jermyn2023}. The MESA EOS is a blend of the OPAL \citep{Rogers2002}, SCVH \citep{Saumon1995}, FreeEOS \citep{Irwin2004}, HELM \citep{Timmes2000}, PC \citep{Potekhin2010}, and Skye \citep{Jermyn2021} EOSes. Radiative opacities are primarily from OPAL \citep{Iglesias1993, Iglesias1996}, with low-temperature data from \citet{Ferguson2005} and the high-temperature, Compton-scattering dominated regime by \citet{Poutanen2017}.  Electron conduction opacities are from \citet{Cassisi2007} and \citet{Blouin2020}. Nuclear reaction rates are from JINA REACLIB \citep{Cyburt2010}, NACRE \citep{Angulo1999} and additional tabulated weak reaction rates \citet{Fuller1985, Oda1994, Langanke2000}.  Screening is included via the prescription of \citet{Chugunov2007}. Thermal neutrino loss rates are from \citet{Itoh1996}.

\section*{Data Availability}
\label{sec:dataAv}
The \kepler data underlying this article are available at the MAST Portal (Barbara A. Mikulski Archive for Space Telescopes), at \url{https://mast.stsci.edu/portal/Mashup/Clients/Mast/Portal.html}.
MESA and Gyre inlists, together with the final lightcurve, are available under this DOI: 10.5281/zenodo.18158471.



\ifarxiv
    \input{output.bbl} 
\else
    \bibliographystyle{mnras}
    \bibliography{bib} 
\fi



\appendix
\section{Isochrones} 
\label{sec:isochrone}
\begin{figure}
  \includegraphics[width=\linewidth]{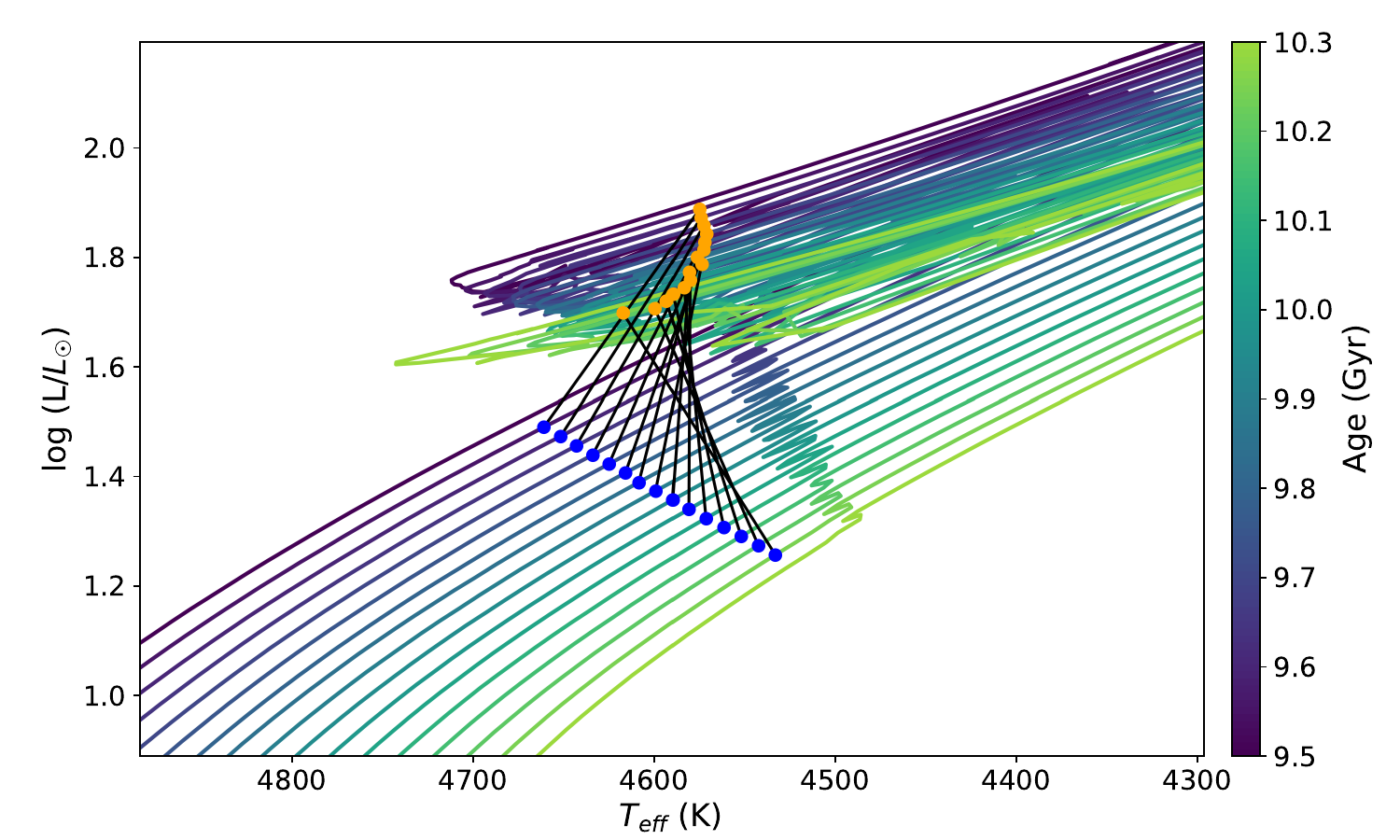}
  \caption{Isochrones from MIST to compare the temperature difference of the two components of \mystar. Blue dots correspond to the RGB and orange dots to the RC star. Models from the same isochrone are connected by a black line. The colour corresponds to the age of the isochrone as shown in the colorbar.}
  \label{fig:isochrones}
\end{figure}

Through our spectral analysis, the effective temperature of the secondary was not obtainable, but is an essential parameter when it comes to the evolution of stars. Because of early test runs of stellar models and estimations via asteroseismic scaling relations, we suspected that both stars have similar temperatures. To further support our assumptions, we downloaded several isochrones from MIST \citep[Mesa Isochrones and Stellar Tracks;][]{Dotter2016,Choi2016} with the closest available metallicity of our system (solar metallicity). Next, we interpolated the isochrones over the masses of the stars and for each isochrone selected the stars closest to our estimated \numax on the RC and RGB, respectively. We limit the shown models to those with masses between 0.8 to 1.4 \Msun and within 10 per cent of the observed \dnu. We compared their effective temperatures as shown in \autoref{fig:isochrones}. Even though we allowed a generous range of models, the biggest temperature difference is in any case never larger than $100 K$.

\section{Correction factor \texorpdfstring{\lowercase{$f$}$_{\Delta\nu}$}{fdnu}}\label{sec:fdnu}
The method described by \citet[][]{LiY2023} to estimate \fdnu accounts for deviations from the $\Delta \nu \propto \sqrt{\rho}$ scaling relation and also includes the surface effect on \dnu. Previous studies often neglected this effect \citep[e.g.][]{Sharma2016,Guggenberger2016}, which can introduce systematic errors \citep{Kjeldsen2008}. To extend this method to RC stars, we employed an updated model grid (\citealt{Chiu2025}; Chiu et al., in prep.) and followed the same calibration procedure as in \citet{LiY2023} to determine the surface effect, using the oscillation frequencies extracted by \citet{Kallinger2019}. We fitted a formula of \fdnu with respect to stellar properties fitted as 
\begin{equation}\label{eq:fDnu}
\begin{aligned}
    f_{\Delta\nu} = \beta_0 &+  \beta_1\log_{10}(\nu_{\rm max}/3090\ \mu{\rm Hz}) \\
    &+\beta_2\log_{10}(\Delta\nu/135.1\ \mu{\rm Hz}) \\
    &+\beta_3(T_{\rm eff}/5777{\ \rm K}) \\
    &+\beta_4(T_{\rm eff}/5777{\ \rm K})^2 \\
    &+\beta_5(T_{\rm eff}/5777{\ \rm K})^3 \\
    &+\beta_6{\rm [M/H]}, \\ 
\end{aligned}
\end{equation}
We obtain the following $\beta$ values for RGB models 
$\beta_{\mathrm{RGB}}=\{3.4218, 0.1953, -0.2381, -7.6958, 7.6003, -2.2785, 0.0234\}$ 
and RC models $\beta_{\mathrm{RC}}=\{-1.717, 0.083, -0.088, 10.842, -14.185, 6.118, 0.028\}$
When applying the method, we observe a systematic offset of about 0.1\,\Msun towards higher masses compared to the description from \citet[][]{Sharma2016}. 
\citet{Pinsonneault2025} also examined systematic uncertainties in computing \fdnu{} and reported differences of up to 3\% in the clump phase, depending on the adopted model physics. Together, this seems to indicate that the internal structure of clump stars is the dominant source of uncertainty in the \fdnu{} determination, which is also supported by our findings.

It is unclear how this offset in mass is related to our modelling result of \mystar. When we do not implement the \epsp offset, our model results clearly overestimate the mass of the RC component (see \autoref{fig:probdis_bad}). Stricter constraints on L, \MH and \Teff or the surface correction \dnusurf itself could reduce this mass offset using this large-scale method, but it nevertheless supports our conclusion that a further investigation into RC models is needed and that results involving RC models are not to be trusted unconditionally. 

\section{Corner plots} 
\label{sec:cornerplots}
\begin{figure*}
  \includegraphics[width=\linewidth,trim={0 0.7cm 0 1.7cm},clip]{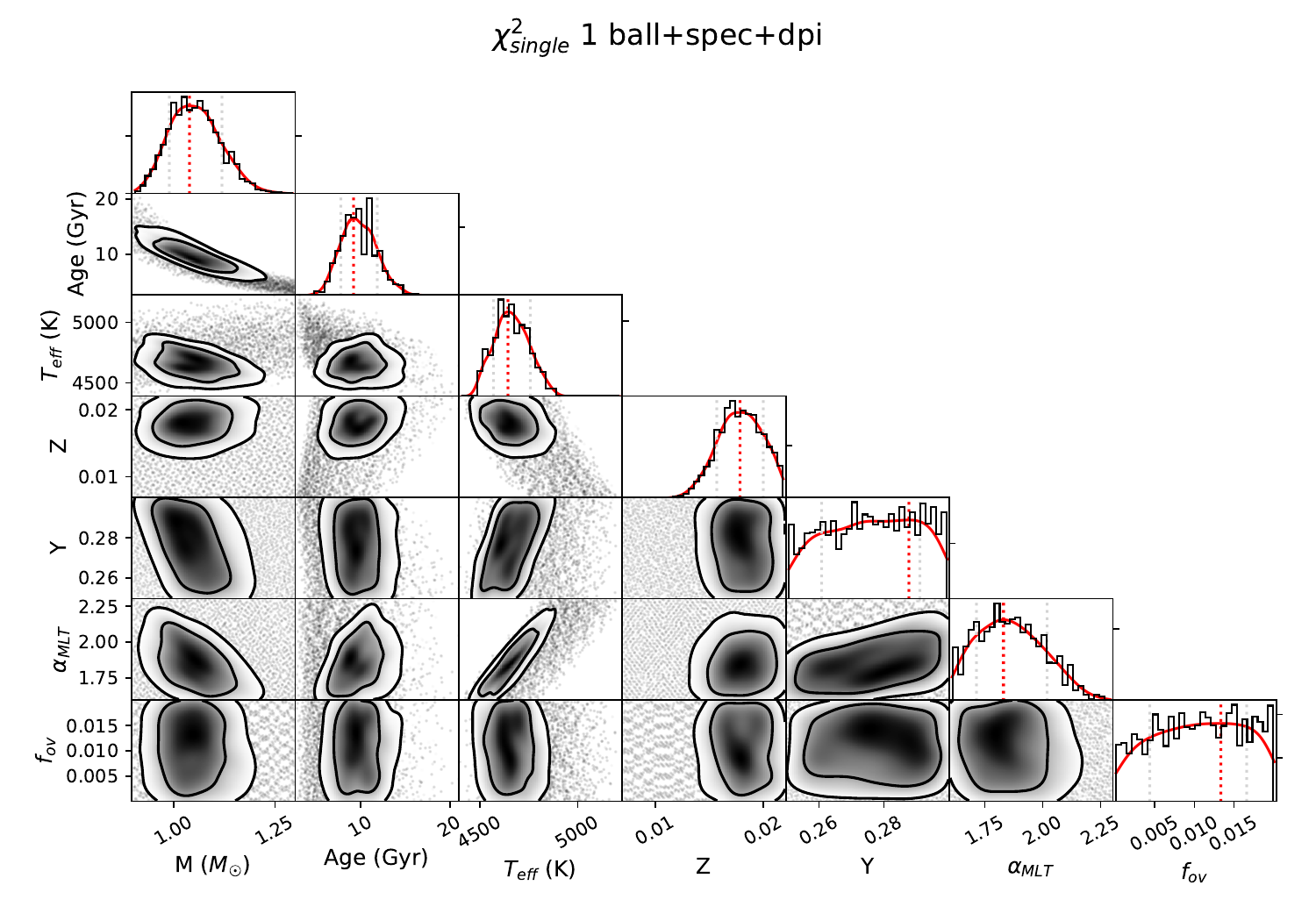}
  \includegraphics[width=\linewidth,trim={0 0.7cm 0 1.7cm},clip]{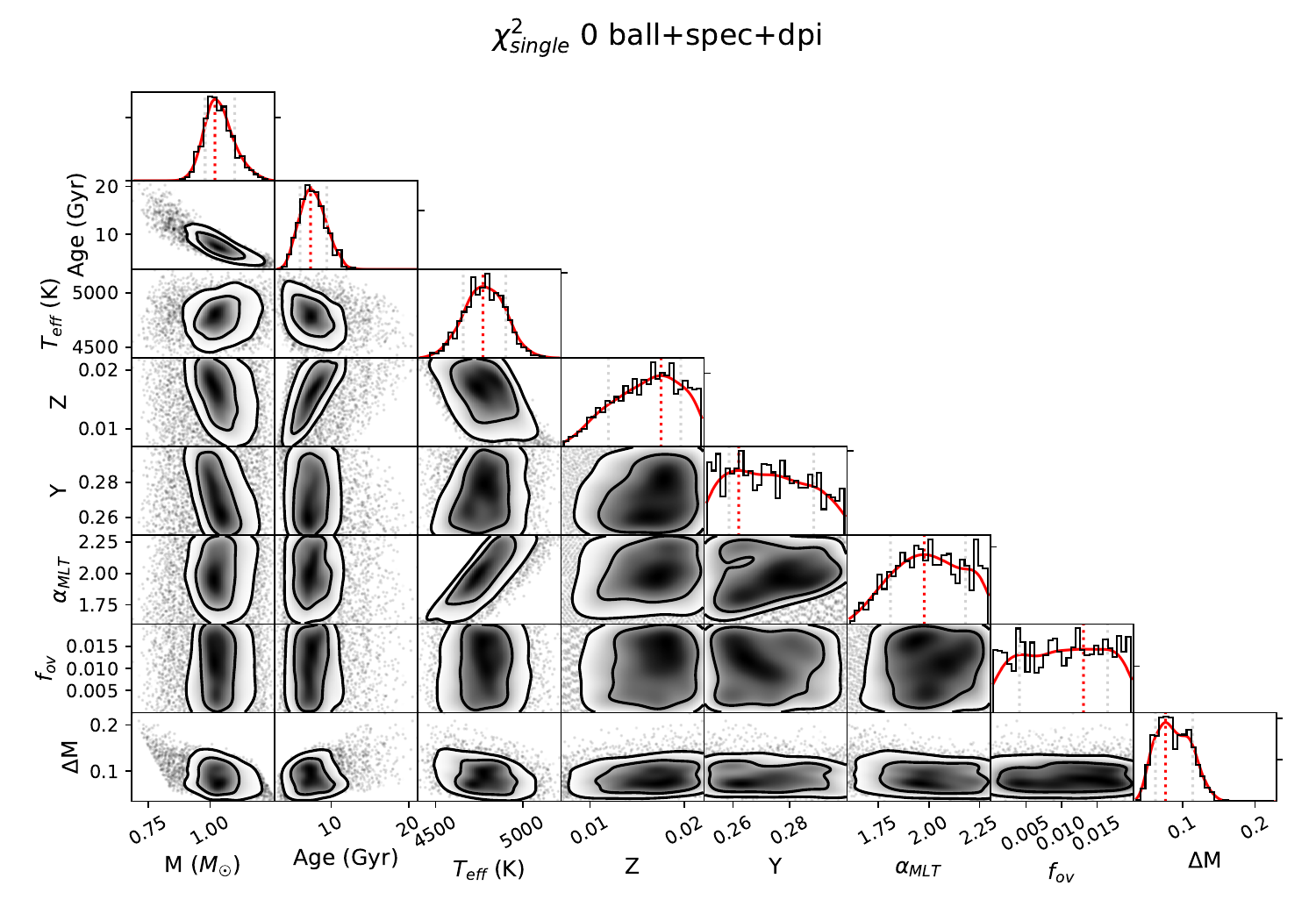}
  \caption{Corner plots for the single-star solutions of the RGB (top) and RC component (bottom) of \mystar. The surface term was less strictly constrained (see \autoref{sec:epsilonOffset}). The grayscale panels show the normalised probability distribution calculated using a kernel density estimation (KDE). The contour lines represent the $1$ and $2\sigma$ confidence regions. Outside the $2\sigma$ level, individual model grid points are shown. The top panel of each column shows the normalised probability distribution in black and the corresponding KDE in red. The mode of the KDE is marked with a red dotted line and the grey dotted lines represent the $1\sigma$ uncertainty.}
  \label{fig:cornerplot}
\end{figure*}

In \autoref{fig:cornerplot} we show the corner plots of the single-star solutions of KIC 10841730, allowing for the $\epsilon$-offset as described in \autoref{sec:epsilonOffset}. 

\bsp	
\label{lastpage}
\end{document}